\documentclass{article}
\usepackage{graphicx}  
\usepackage{amsmath}   
\usepackage{amssymb}   
\usepackage{bm} 
\usepackage{dcolumn}
\usepackage{color}
\usepackage{mathrsfs}
\usepackage{amsfonts}
\usepackage{varioref}
\usepackage[margin=1.37in]{geometry}
\usepackage{slashed}
\RequirePackage[colorlinks,citecolor=blue,urlcolor=magenta,linkcolor=blue]{hyperref}
\def\p{\partial}
\def\e{\epsilon}

\def\be{\begin{equation}}
\def\ee{\end{equation}}

\labelformat{section}{Section #1} 
\labelformat{subsection}{Section #1} 
\labelformat{subsubsection}{Section #1}
\labelformat{subsubsubsection}{Section #1}
\labelformat{equation}{Eq.~(#1)} 
\labelformat{figure}{Fig.~#1} 
\labelformat{subfigure}{Fig.~\thefigure#1} 
\labelformat{table}{Tab.~#1} 
\labelformat{appendix}{Appendix #1}

\title{\bf Effective action for $\phi^4$-Yukawa theory via 2PI formalism in the inflationary de Sitter spacetime }
\author{Sourav Bhattacharya\footnote{sbhatta.physics@jadavpuruniversity.in}\,\, and Kinsuk Roy\footnote{kinsukr.physics.rs@jadavpuruniversity.in}\\
\small{Relativity and Cosmology Research Centre, Department of Physics, Jadavpur University, Kolkata 700 032, India}\\}
\begin{document}
\maketitle
\begin{abstract}
\noindent
We consider a scalar field theory with quartic self interaction, Yukawa coupled to fermions  in the inflationary de Sitter spacetime background. The scalar has a classical background plus quantum fluctuations, whereas the fermions are taken to be quantum. We derive for this system the effective action and the effective potential via the two particle irreducible (2PI)  formalism. This formalism provides an opportunity to find out resummed or  non-perturbative expressions for some series of  diagrams. We have considered the two loop vacuum graphs and have computed the local part of the effective action. The various resummed counterterms corresponding to self energies, vertex functions and the tadpole have been explicitly found out. The  variation of the renormalised effective potential for massless fields has been investigated numerically. We show that for the potential to be bounded from below, we must have $\lambda \gtrsim 16 g^2$, where $\lambda$ and $g$ are respectively the quartic and Yukawa couplings.  We emphasise the qualitative differences of this non-perturbative calculation with that of the standard 1PI perturbative ones in de Sitter. The qualitative differences of our result with that of the flat spacetime has also been pointed out.
\end{abstract}
\vskip .5cm

\newpage

\tableofcontents

\section{Introduction}\label{S1}
 
The standard hot big bang model of cosmology  has been greatly successful in explaining  the redshift of light coming from  distant galaxies, the origin of the cosmic microwave background radiation, abundance of light elements, the  large scale cosmic structure formation etc.. However, this model cannot explain a few puzzling features of our universe, such as its spatial flatness, the horizon problem associated with the spatial isotropy at very large scales, and the hitherto unobserved relics such as the magnetic monopoles. The epoch of primordial cosmic inflation is a proposed phase of a very rapid, near exponential accelerated expansion of our very early universe.  Such inflationary phase not only gives answers to these                                                                                                                                                                                                                                                                                                                                                                                                                                                                                                                                                                                                                                                                                                                                                                                                                                                                                                                                                                                                                                                                                                                                                                                                                                                                                                                                                                                                                                                                                                                                                                                                                                                                                                                                                                                                                                                                                                                                                                                                                                                                                                                                                                                                                                                                                                                                                                                                                                                                                                                                                                                                                                                                                                                                                                                                                                                                                                                                                                                                                                                                                                                                                                                                                                                                                                                                                                                                                                                                                                                                                                                                                                                                                                                                                                                                                                                                                                                                                                                                                                                                                                                                                                                                                                                                                                                                                                                                                                                                                                                                                                                                                                                                                                                                                                                                                                                                                                                                                                                                                                                                                                                                                                                                                                                                                                                                                                                                                                                                                                                                                                                                                                                                                                                                                                                                                                                                                                                                                                                                                                                                                                                                                                                                                                                                                                                                                                                                                                                                                                                                                                                                                                                                                                                                                                                                                                                                                                                                                                                                                                                                                                                                                                                                                                                                                                                                                                                                                                                                                                                                                                                                                                                                                                                                                                                                                                                                                                                                                                                                                                                                                                                                                                                                                                                                                                                                                                                                                                                                                                                                                                                                                                                                                                                                                                                                                                                                                                                                                                                                                                                                                                                                                                                                                                                                                                                                                                                                                                                                                                                                                                                                                                                                                                                                                                                                                                                                                                                                                                                                                                                            three problems, but also provides a suitable framework to generate primordial quantum  density perturbations and correlation functions, as  seeds to the large scale cosmic structures we observe today in the sky, see~\cite{Wein} and references therein for various theoretical and observational aspects of cosmic inflation. 
 
 The inflation requires an exotic matter field with negative isotropic pressure. Traditionally, this is explained by a  scalar field, called inflaton, slowly moving down a potential. Also, after sufficient number of $e$-foldings, the universe  must gracefully exit the inflationary period. This graceful exit may also specify the current observed value of the cosmological constant. It turns out that only $10\%$ change to this value would modify the evolution of our universe greatly, known as the cosmic coincidence  problem. We refer our reader to e.g.~\cite{Tsamis, Inagaki:2005qp, Ringeval, Padmanabhan, Alberte, Appleby, Evnin:2018zeo} and references therein for various attempts to address this issue.

The de Sitter (\ref{y0}) or quasi de Sitter spacetime is believed to be the metric appropriate for the inflationary phase. For such time dependent background, understanding quantum fluctuations is an important task. We refer our reader to~\cite{Chernikov:1968zm, Bunch:1978yq, Linde:1982uu, Starobinsky:1982ee, Allen:1985ux, Allen} for discussion on field quantisation in de Sitter background. The dynamics of a light scalar field can be very non-trivial in such a background. In particular, a massless but non-conformally invariant (such as a massless minimally coupled scalar, gravitons) field cannot have a de Sitter invariant Wightman function~\cite{Allen:1985ux, Allen}. This leads to  the appearance of logarithm of the scale factor in quantum amplitudes and hence breakdown of the perturbative  expansion at late times, known as the secular effect~\cite{Floratos}. Such large logarithms are chiefly related to the super-Hubble, deep infrared fluctuations at late times. They can lead to dynamical generation of the field  rest mass. For various aspects of this non-perturbative effect including mass generation, cosmic decoherence and entanglement, we refer our reader to e.g.~\cite{davis, Onemli:2002hr, Brunier:2004sb, Prokopec:2003tm, Kahya:2009sz, Miao:2012bj, Boyanovsky:2012qs, Onemli:2015pma, Prokopec:2007ak, Cabrer:2007xm, Akhmedov:2011pj, Akhmedov:2012pa, Beneke:2012kn, Akhmedov:2013xka, Boran:2017fsx, Karakaya:2017evp, Liao:2018sci, Akhmedov:2019cfd, Glavan:2019uni, Boyanovski, Bhattacharyya:2024duw, Brahma:2024ycc} and references therein.
Resummation of these secular logarithms has  been attempted in numerous occassions, see e.g.~\cite{Burgess:2009bs, Burgess:2015ajz, Serreau:2013eoa, Moreau:2018lmz,  Moreau:2018ena, Youssef:2013by, Ferreira:2017ogo, Kitamoto:2018dek, Baumgart:2019clc, Kamenshchik:2020yyn, Kamenshchik:2021tjh, Kamenshchik:2024ybm, Bhattacharya:2022wjl, Bhattacharya:2023xvd, Miao:2021gic, Miao2} and references therein. In particular, for scalar  field theories with non-derivative interaction, the late time stochastic formalism is an excellent way to do resummation~\cite{Starobinsky:1994bd} (see also e.g.~\cite{Finelli:2008zg, Garbrecht:2013coa, Cho:2015pwa, Ebadi:2023xhq, Cruces:2022imf} and references therein). However, the same issue for field theories with derivative interactions such as gravity, remains largely as an open question to this date. \\

\noindent
The effective action technique can be a very strong tool to address the early inflationary quantum fluctuations. In this work we wish to derive the effective action for a $\phi^4$-Yukawa theory in the de Sitter background by using the non-perturbative,  2 particle irreducible (2PI) (see e.g.~\cite{Berges:2004yj} for a vast review) formalism. Earlier works on renormalisation and resummation, self energy computation, fermion mass generation and decoherence for the Yukawa theory in de Sitter can be seen in~\cite{Prokopec:2003qd, Duffy:2005ue, Garbrecht, Bhattacharya:2023twz, Bhattacharya:2023yhx}. See \cite{Miao:2006pn} for one loop effective action of Yukawa theory in de Sitter and also~\cite{Miao:2020zeh} for inclusion of gauge coupling. Effective action and correlation function for the Yukawa theory using the influence functional technique can be seen 
in~\cite{Boyanovsky:2016exa}. For a renormalisation group improvement of the ultraviolet  (UV) or high field limit of the one loop effective action for the $\phi^4$-Yukawa theory in de Sitter can be seen in~\cite{Hardwick:2019uex}. We also refer our reader to~\cite{Bhattacharya:2025eqf} for a study on the effect of Yukawa coupling on cosmological correlation functions. We further refer our reader to \cite{Toms:2018wpy, Toms:2018oal, Inagaki:1993ya,  Elizalde:1995bm, Inagaki:2015eza},  for discussion on effective action with Yukawa and four fermion interactions in  general curved spacetimes using the Schwinger-DeWitt local expansion technique. This technique is UV effective, and hence cannot probe physics at scales  comparable to the Hubble horizon.\\
 
\noindent
Earlier, the 2PI  technique was used in de Sitter to compute the non-perturbative effective action for $\phi^4$ or $O(N)$ scalar field theory at two loop Hartree approximation in~\cite{Arai:2012sh, Arai:2013jna, LopezNacir:2013alw}. We will focus only on the local part of the effective action in this work. In particular, we will see below the renormalisation in the $\phi^4$-Yukawa theory has  non-trivial features, which is qualitatively very different from that of the standard perturbative approach. We will also emphasise in the due course, the non-trivial results such non-perturbative technique may bring in,  in the de Sitter background. We will also emphasise the qualitative differences of our result with that of the flat spacetime.\\

 \noindent
 The rest of the paper is organised is as follows. In the next section we sketch the basic technical framework we will be working in. In \ref{S31}, following~\cite{Arai:2012sh, Arai:2013jna, LopezNacir:2013alw} we briefly discuss the derivation of the two loop 2PI effective action for $\phi^4$ theory in the Hartree approximation. Here we also derive the two and three loop vacuum graphs (both local and non-local parts) for a massless and minimal scalar without any background field, in the leading power of the secular logarithm, using the Schwinger-Keldysh formalism (e.g.~\cite{Hu, Adshead}, and references therein). Next in \ref{S32}, we  compute the local part of the 2PI effective action for the $\phi^4$-Yukawa theory at two loop. Finally we conclude in \ref{S4}. The four appendices contain some technical detail of the calculations. We will work with the mostly positive signature of the metric in $d = 4 - \epsilon$ ($\epsilon = 0^{+}$) dimensions and will set $c = 1 = \hbar$ throughout.  Also for the sake of brevity and to save space, we will denote for powers of propagators and logarithms respectively as, $(i\Delta)^n\equiv i\Delta^n$ and also $(\ln a )^n \equiv \ln^n a$.

\section{The basic set up}\label{S2}

We wish to briefly review below the basic framework we will be working in, following chiefly~\cite{Onemli:2002hr, Brunier:2004sb, Miao:2006pn}. 
The metric for the inflationary de Sitter spacetime reads respectively in the cosmological and conformal temporal coordinates
\begin{eqnarray}
ds^2 = -dt^2 + a^2(t) d{\vec x}^2= a^2(\eta) \left(-d\eta^2 +  d{\vec x}^2\right)
\label{y0}
\end{eqnarray}
where $a(t)=e^{Ht}$ or $a(\eta)=  -1/H\eta$ is the de Sitter scale factor and $H=\sqrt{\Lambda/3}$ is the de Sitter Hubble rate, with $\Lambda$ being the cosmological constant.  We have the range $0 \leq t \ensuremath{<} \infty$, so that $-H^{-1}\leq \eta \ensuremath{<}0^-$. Note that any temporal level of the initial hypersurface can 
be achieved as we wish, by exploiting the time translation symmetry (along with a scaling of the spatial coordinates) of de Sitter.\\ 

\noindent
The bare action corresponding to the matter sector reads,
\begin{eqnarray}
S= \int a^d d^d x \left[ -\frac12 g^{\mu\nu} (\nabla_{\mu} \Phi')(\nabla_{\nu} \Phi') -\frac12 m_0^2 \Phi'^2 - \frac{\lambda_0}{4!} \Phi'^4 - \frac{\beta_0}{3!} \Phi'^3 - \tau_0 \Phi' + i\overline{\Psi}\slashed{\nabla} \Psi - M_{0} \overline{\Psi} \Psi -g_0 \overline{\Psi} \Psi \Phi'   \right]
\label{y1}
\end{eqnarray}
where $\slashed{\nabla}= \tilde{\gamma}^{\mu}\nabla_{\mu}$, and $\tilde{\gamma}^{\mu}$ and $\nabla_{\mu}$ are respectively the curved space gamma matrices and the spin covariant derivative. Also, since we are working with the mostly positive signature of the metric, we will take the anti-commutation relation for the $\gamma$-matrices as
\be
[\tilde{\gamma}_{\mu},\tilde{\gamma}_{\nu}]_+= -2 g_{\mu\nu} \, {\bf I}_{d\times d}=-2 a^2\eta_{\mu\nu} \, {\bf I}_{d\times d}
\label{Ac}
\ee
Thus we may choose, $\tilde{\gamma}_{\mu} = a(\eta)\gamma_{\mu}$, where $\gamma_{\mu}$'s are the flat space gamma matrices.
Defining the field strength renormalization, $\varphi=\Phi'/\sqrt{Z}$ and $\psi=\Psi/\sqrt{Z_f}$, we have
\begin{eqnarray}
&&S= \int a^d d^d x\left[ -\frac{Z}{2} \eta^{\mu\nu} a^{-2} (\p_{\mu} \varphi)(\p_{\nu} \varphi) -\frac{1}{2} Z m^2 \varphi^2  - \frac{Z^2 \lambda_0}{4!} \varphi^4  - \frac{\beta_0 Z^{3/2}}{3!} \varphi^3  - \tau \sqrt{Z} \varphi \right. \nonumber\\ && \left.- i Z_f\overline{\psi}\slashed{\nabla} \psi - M Z_f  \overline{\psi} \psi  -g_0 Z_f \sqrt{Z}\, \overline{\psi} \psi \varphi\right] 
\label{y2}
\end{eqnarray}
 We next write
\begin{eqnarray}
&&Z= 1+\delta Z \qquad Z m^2 = m_0^2+\delta m^2 \qquad Z^2 \lambda_0 = \lambda +\delta \lambda \qquad \beta_0 Z^{3/2}= \delta \beta   \nonumber\\ && \tau \sqrt{Z} =\delta \tau \qquad 
Z_f =1+\delta Z_f \qquad M Z_f= M_0+\delta M \qquad g_0 Z_{f}\sqrt{Z}= g+\delta g  
\label{y3}
\end{eqnarray}
The above decomposition splits \ref{y2} as
\begin{eqnarray}
&&S= \int a^d d^d x\left[  -\frac{1}{2} \eta^{\mu\nu} a^{-2} (\p_{\mu} \varphi)(\p_{\nu} \varphi)-\frac12 m_0^2\varphi^2  - \frac{\lambda}{4!} \varphi^4    - \delta\tau \varphi  + i \overline{\psi}\slashed{\nabla} \psi - M_0 \overline{\psi}\psi   -g \overline{\psi} \psi \varphi \right.\nonumber\\&&\left.-\frac{\delta Z}{2} \eta^{\mu\nu} a^{-2} (\p_{\mu} \varphi)(\p_{\nu} \varphi)  - \frac12 \delta m^2 \varphi^2-\frac{\delta \lambda}{4!} \varphi^4 - \frac{\delta\beta}{3!} \varphi^3   + i \delta Z_f \overline{\psi}\slashed{\nabla} \psi  -\delta M \overline{\psi}\psi  - \delta g \overline{\psi} \psi \varphi\right]
\label{y4}
\end{eqnarray}

\noindent
Let us now come to the issue of the scalar field. The exact propagator for a massive scalar can be seen in \ref{d1} of \ref{D}. Although we will chiefly work with a massive scalar in this paper (note that a background field makes the scalar effectively massive, even if the scalar has zero rest mass), we also wish to additionally compute some vacuum graphs for a massless and minimally coupled scalar field (\ref{A}), which is of special interest in de Sitter.  
The relevant propagator such a  scalar reads~\cite{Onemli:2002hr},
\be
i\Delta(x,x')= A(x,x')+ B(x,x')+ C(x,x')
\label{props1}
\ee
where 
\begin{eqnarray}
&&A(x,x') = \frac{H^{2-\e} \Gamma(1-\e/2)}{4\pi^{2-\e/2}}\frac{1}{y^{1-\e/2}} = \frac{\Gamma(1-\e/2)}{2^2 \pi^{2-\e/2}}\frac{(aa')^{-1+\e/2}}{(\Delta x)^{2-\e}} \nonumber\\
&&B(x,x') =  \frac{H^{2-\e} \Gamma(3-\e)}{2^{4-\e}\pi^{2-\e/2}\Gamma(2-\e/2)}\left[-\frac{2\Gamma(3-\e/2) \Gamma(2-\e/2)(aa'H^2/4)^{\e/2} }{\e \Gamma(3-\e)}  (\Delta x^2)^{\e/2}+  \left(\ln (aa')+ \frac{2}{\e}\right)  \right]\nonumber\\
&&C(x,x')= \frac{H^{2-\e} }{(4\pi)^{2-\e/2}} \sum_{n=1}^{\infty} \left[\frac{\Gamma(3-\e+n)}{n\Gamma(2-\e/2+n)}\left(\frac{y}{4} \right)^n- \frac{\Gamma(3-\e/2+n)}{(n+\e/2)\Gamma(2+n)}\left(\frac{y}{4} \right)^{n+\e/2}  \right]
\label{props2}
\end{eqnarray}
The de Sitter invariant biscalar interval reads
\be
y(x,x')= aa'H^2 \Delta x^2 = a(\eta)a(\eta')H^2 \left[ |\vec{x}-\vec{x}'|^2- (\eta-\eta')^2\right]
\label{props3}
\ee
The $\ln (aa')$ term appearing in $B(x,x')$  in \ref{props2} makes the propagator non-invariant under de Sitter symmetry transformations. This is a unique feature of an exactly massless and minimal scalar field theory in de Sitter. In other words, a smooth $m^2\to 0$ limit for a scalar in de Sitter may not even exist. 

There are four propagators pertaining to the in-in or the Schwinger-Keldysh formalism one needs to use in a cosmological framework e.g.~\cite{Hu} (\ref{A}), characterised by suitable four complex distance functions, $\Delta x^2$,
\begin{eqnarray}
&&\Delta x^2_{++} =\left[ |\vec{x}-\vec{x'}|^2- (|\eta-\eta'|-i\e)^2\right]= (\Delta x^2_{--})^{*}\nonumber\\
&&\Delta x^2_{+-} =\left[ |\vec{x}-\vec{x'}|^2- ((\eta-\eta')+i\e)^2\right]= (\Delta x^2_{-+})^{*} \qquad (\e=0^+)
\label{props3'}
\end{eqnarray}
The first two correspond respectively to the Feynman and anti-Feynman propagators, whereas the last two correspond to the two Wightman functions. 
From \ref{props2}, we have in the coincidence limit for all the four propagators
\be 
i\Delta(x,x) = \frac{H^{2-\e} \Gamma(2-\e)}{2^{2-\e} \pi^{2-\e/2} \Gamma(1-\e/2)}\left(\frac{1}{\e}+\ln a  \right)
\label{y6}
\ee

Using the above expression, we may compute, for example, the one loop self energy bubble corresponding to the quartic self interaction diagrams.   The corresponding  one loop  mass renormalisation counterterm reads~\cite{Brunier:2004sb}
\be 
\delta m_{\lambda}^2 = - \frac{\lambda H^{2-\e} \Gamma(2-\e)}{2^{3-\e} \pi^{2-\e/2} \Gamma(1-\e/2)\e}
\label{y7}
\ee
%

\section{Computation of the local effective action}\label{S3}
\subsection{Warm up -- a scalar with $\phi^4$ self-interaction}\label{S31}
\begin{figure}[h!]
\begin{center}
  \includegraphics[width=10.0cm]{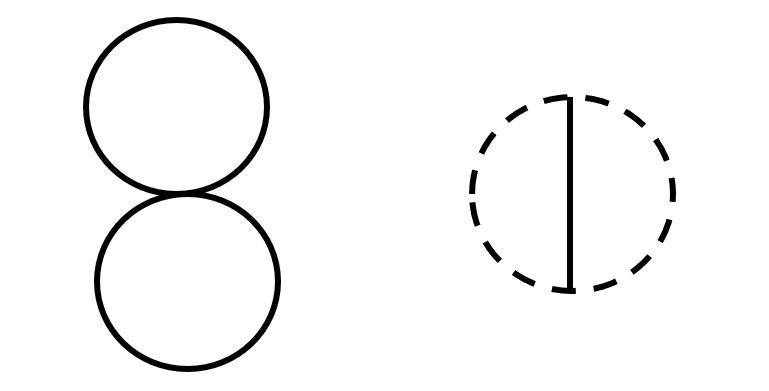}
 \caption{ \it \small (Left)  The two loop (lowest order) 2PI vacuum diagram for the quartic self interaction. (Right) The two loop (lowest order) 2PI diagram for the Yukawa interaction. Solid line stands for scalar, whereas a dashed line stands for fermion. The propagators are exact here.}
  \label{f0}
\end{center}
\end{figure}

The computation of the local, non-perturbative effective action in Hartree approximation in de Sitter was done earlier in~\cite{Arai:2012sh, Arai:2013jna, LopezNacir:2013alw}. For our future purpose, we wish to briefly sketch below the outline of that computation in our own notation.
We begin with the action relevant for our purpose
\begin{eqnarray}
&&S[\varphi]= -\int a^d d^d x \left[\frac12 (\nabla_{\mu} \varphi)(\nabla^{\mu} \varphi) + \frac12( m_0^2+\delta m^2) \varphi^2 +  \frac{(\lambda+\delta \lambda) \varphi^4}{4!} +  \delta \xi R \varphi^2 \right] 
\label{y13'}
\end{eqnarray}
We decompose the field as
$$\varphi = v +\phi$$
where $v$ is the background field and $\phi$ is the quantum fluctuation.  The corresponding 2PI effective action reads~\cite{Berges:2004yj}
\begin{eqnarray}
\Gamma_{\rm 2PI}[v, iG] = S[v] - \frac{i}{2} \ln {\rm det}\ i G (x,x') +\frac{1}{2} {\rm Tr} \int (aa')^d d^dx d^d x' i \Delta^{-1}(x,x') iG(x',x) + i\Gamma_2 [v,iG]
\label{y13}
\end{eqnarray}
where  $i\Delta(x,x')$ is the free whereas $iG(x,x')$ is the exact propagator for the scalar.    The above expression explicitly reads for our theory
\begin{eqnarray}
&&\Gamma_{\rm 2PI}[v, iG] = -\int a^d d^d x \left[\frac12 (\nabla_{\mu} v)(\nabla^{\mu} v) + \frac12 (m_0^2+\delta  m_{1}^2) v^2 + \frac{\lambda_{1} v^4}{4!} + \delta \xi_{1} R v^2 \right]\nonumber\\&& - \frac{i}{2} \ln {\rm det}\ i G (x,x) +\frac{1}{2} \int a^d d^dx \left[\Box - (m_0^2+\delta m^2_{2} )- 2\delta \xi_{2} R- \frac{\lambda_{2} v^2}{2}  \right]  iG(x,x) + i\Gamma_2 [v,iG] 
\label{y14}
\end{eqnarray}
$i\Gamma_2[v,iG]$ generates the 2PI vacuum graphs, and the vertex counterterms are contained within $\lambda_1$, $\lambda_2$. At the leading order and at two loop, we have
\begin{eqnarray}
&&i\Gamma_{2}[v, iG] = - \frac{\lambda_{3}}{2^3} \int a^d d^d x \ i G^2(x,x) 
\label{y15}
\end{eqnarray}
Note that the above double bubble vacuum graph is purely local (the first of \ref{f0}). This is  called the Hartree approximation. We also have 
$$\lambda_i= \lambda+\delta \lambda_i\qquad i=1,2,3$$
Let us consider the  equation of motion satisfied the Green function found from \ref{y14},   
\begin{eqnarray}
&&  \left[\Box -(m_0^2+ \delta m^2_{2}) - 2\delta \xi_{2} R- \frac{\lambda_{2} v^2}{2}\right] i G(x,x')=  \frac{i\delta^d (x-x')}{a^d} - 2 \int a''^d d^d x'' \frac{\delta i \Gamma_2}{\delta iG(x,x'')}iG(x'',x')  
\label{y17}
\end{eqnarray}
where we have used
$$\int d^d x'' a''^d\ i G(x,x'')iG^{-1}(x'',x') = \frac{\delta^d(x-x')}{a^d}$$
The term appearing in the last term on the right hand side of \ref{y17}, $2\delta i \Gamma_2/\delta iG(x,x')$ is basically $-i$ times the ${\cal O}(\lambda)$ 1PI self energy. This whole last term explicitly reads
\begin{eqnarray}
&& \frac{\lambda_3}{2} iG(x,x) iG(x,x') 
\label{y18}
\end{eqnarray}
Since $iG(x,x)$ is purely local, \ref{y17} gives us an opportunity to resum it easily. The resummed local self energy dynamically generates a rest mass at late times, say $m_{\rm dyn}^2$. 
Accordingly, we assume that \ref{y17}  at late times takes the form
\begin{eqnarray}
&&  \left[\Box - \frac{\lambda v^2}{2}- m_{\rm dyn}^2\right] i G(x,x')=  \frac{i\delta^d (x-x')}{a^d} 
\label{y19}
\end{eqnarray}

\noindent
Let us now come to the non-perturbative renormalisation scheme. From \ref{d60} (\ref{D}) valid for an arbitrary massive scalar, we write for the sake of brevity
\be
iG(x,x) = m^2_{\rm dyn, eff} f_d + H^2 f'_d + f_{\rm fin}
\label{y41a1}
\ee
where
\begin{eqnarray}
&&f_d = -\frac{H^{-\e}}{2^{3-\e}\pi^{2-\e/2}\e} \qquad  f'_d = -\frac{H^{-\e}}{2^{2-\e}\pi^{2-\e/2}\e} \left(1- \frac{\gamma \e}{2} \right) \nonumber\\
&& f_{\rm fin} = \frac{H^2}{2^3\pi^2} \left( 1- \frac{1}{2} \frac{m^2_{\rm dyn, eff}}{H^2}  \right) \left[ \frac{1}{s} -\frac{2}{1-s}  -\left( \psi(1+s) +\psi (1-s)\right)  \right],
\label{y41a2}
\end{eqnarray}
where $\psi$ is the digamma function, and we have abbreviated,
\be
s 
= \frac32 - \left( \frac94-\frac{{ m}_{\rm dyn, eff}^2}{H^2}\right)^{1/2}
\label{y57'}
\ee
From  the field equation for the propagator, \ref{y17}, \ref{y19}, we write
\begin{eqnarray}
m^2_{\rm dyn, eff} = m_0^2 +\delta m_2^2 +\frac{(\lambda +\delta \lambda_2) v^2}{2} + \frac{\lambda +\delta \lambda_3}{2} iG(x,x)+2\delta \xi_2 R
\label{y41a3}
\end{eqnarray}
where we have defined
\be
m^2_{\rm dyn, eff} =m_0^2 +\frac{\lambda f_{\rm fin}}{2} + \frac{\lambda v^2}{2} 
\label{y41a4}
\ee
We next plug in \ref{y41a1} into \ref{y41a3}.  We first infer that
\be
\delta \lambda_2=\delta \lambda_3, \qquad \delta \xi_2 =0 
\label{y41a5}
\ee
We also have the self consistency condition
\be
\delta m_2^2 + \frac12 (\lambda +\delta \lambda_2) (f_d m_0^2 + H^2 f'_d) + \frac{v^2}{2} \left(\delta \lambda_2 + \frac{\lambda}{2}\left(\lambda +\delta \lambda_2\right) f_d    \right) +\frac12 \left(\delta \lambda_2 + \frac{\lambda}{2}\left(\lambda +\delta \lambda_2\right) f_d\right)f_{\rm fin}=0    
\label{y41a5}
\ee
Next by individually setting the coefficients of $v^2$ and $f_{\rm fin}$  to zero, we obtain the non-perturbative renormalisation counterterms
\begin{eqnarray}
\delta \lambda_2 = -\frac{\lambda^2 f_d}{2\left( 1+ \frac{\lambda f_d}{2}\right)}, \qquad \delta  m_2^2 =- \frac{\lambda  (f_d m_0^2 + H^2 f'_d)  }{2\left( 1+ \frac{\lambda f_d}{2}\right)}
\label{y41a36}
\end{eqnarray}
We now write down the effective action as 
\begin{eqnarray}
&&\Gamma_{\rm 2PI}[v, iG] 
 =  -\int a^d d^d x \left[\frac12 (\nabla_{\mu} v)(\nabla^{\mu} v) + \frac12 (m_0^2 +\delta m_1^2) v^2 +\frac{(\lambda+\delta\lambda_1) v^4}{4!}   \right]  -\frac{1}{2} \int a^d d^d x \int d m^2_{\rm dyn, eff}   ( {m}^2_{\rm dyn, eff} f_d \nonumber\\
&&+ H^2 f'_d + f_{\rm fin})+ \frac{\lambda +\delta \lambda_2}{2^3} \int a^d d^d x  \left(m^2_{\rm dyn, eff} f_d + H^2 f'_d + f_{\rm fin}\right)^2
\label{y41a6}
\end{eqnarray}
For the  last two terms in the effective action, we respectively have 
\begin{eqnarray}
    \int d m^2_{\rm dyn,eff}iG(x,x)&=&\frac{1}{2}f_d m^4_{\rm dyn,eff}+H^2m^2_{\rm dyn,eff}f_d'+\int d m^2_{\rm dyn,eff} f_{\rm fin}\nonumber\\&=&\frac{1}{2}m_0^4f_d+v^2. \frac{1}{2}m_0^2\lambda f_d+v^4. \frac{1}{8}\lambda^2f_d+v^2 f_{\rm fin}. \frac{1}{4}\lambda^2f_d+f_{\rm fin}. \frac{1}{2}m_0^2\lambda f_d \nonumber\\&+&f_{\rm fin}^2.\frac{1}{8}\lambda^2f_d +H^2\left(m_0^2+\frac{1}{2}\lambda v^2+\frac{1}{2}\lambda f_{\rm fin}\right)f_d'+\int dm^2_{\rm dyn,eff} f_{\rm fin}
    \label{52eff}
\end{eqnarray}
and
\begin{eqnarray}
   && iG^2(x,x) =m_0^4f_d^2+v^2. m_0^2\lambda f_d^2 +v^4. \frac{1}{4}\lambda^2f_d^2+v^2f_{\rm fin}. \lambda f_d \left(1+\frac{1}{2}\lambda f_d\right)+f_{\rm fin}. 2m_0^2f_d\left(1+\frac{1}{2}\lambda f_d\right)\nonumber\\&& +f_{\rm fin}^2. \left(1+\frac{1}{2}\lambda f_d\right)^2+H^4{f'}_d^2+2H^2f'_d\left[\left(m_0^2+\frac{1}{2}\lambda v^2+\frac{1}{2}\lambda f_{\rm fin}\right)f_d+f_{\rm fin}\right]
    \label{53eff}
\end{eqnarray}
Plugging the two above expressions into \ref{y41a6}, we first conclude that
\be
\delta \lambda_1 = 3\delta \lambda_2,   \qquad \delta m_2^2=\delta m_1^2, \qquad \delta \xi_1^2 =0
\label{CT}
\ee
We now collect and group terms proportional to $v^2$,     $v^4$, $v^2f_{\rm fin}$, $f_{\rm fin}$ and $f_{\rm fin}^2$ in \ref{y41a6}. Using \ref{y41a4}, \ref{y41a5}, \ref{y41a36} and \ref{CT}, we obtain the following respective coefficients  after a little algebra
\begin{eqnarray}
&&v^2 :  \  \frac{1}{8}(\lambda+\delta\lambda_2)m_0^2\lambda f_d^2-\frac{1}{4}m_0^2\lambda f_d-\frac{1}{2}\delta m_2^2+\frac{1}{4}(\lambda+\delta\lambda_2)H^2{f'}_d\left(1+\frac{1}{2}\lambda f_d\right)-\frac{1}{4}H^2\lambda {f'}_d=0 \nonumber\\
 &&    v^4 : \  -\frac{\delta\lambda_2}{8}-\frac{1}{16}\lambda^2f_d+\frac{1}{32}(\lambda+\delta\lambda_2)\lambda^2f_d^2=0\nonumber\\
 &&f_{\rm fin} : \ \frac{1}{8}(\lambda+\delta\lambda_2)2m_0^2f_d\left(1+\frac{1}{2}\lambda f_d\right)-\frac{1}{4}m_0^2\lambda f_d+\frac{1}{4}H^2{f'}_d\left(1+\frac{1}{2}\lambda f_d\right)(\lambda+\delta\lambda_2)-\frac{1}{4}H^2\lambda{f'}_d=0\nonumber\\
   &&  f_{\rm fin}^2 : \ \frac{1}{8}(\lambda+\delta\lambda_2)\left(1+\frac{1}{2}\lambda f_d^2\right)^2-\frac{1}{16}\lambda^2f_d=\frac{\lambda}{8} \nonumber\\&&
     v^2f_{\rm fin} : \ \frac{1}{8}(\lambda+\delta\lambda_2)\lambda f_d\left(1+\frac{1}{2}\lambda f_d\right)-\frac{1}{8}\lambda^2f_d=0
        \label{58eff}
\end{eqnarray}
Putting everything together, the 2PI effective action now reads,
\begin{eqnarray}
  &&  \Gamma_{\rm 2PI}[v, iG]=-\int a^dd^dx\left[\frac12 (\nabla_{\mu} v)(\nabla^{\mu} v)
    +\frac{1}{2}m_0^2v^2+\frac{\lambda }{4!}v^4-\frac{\lambda}{8} f_{\rm fin}^2+\frac{1}{2}\int d{m}^2_{\rm dyn,eff}f_{\rm fin}\right. \nonumber\\&&\left.+\left(\frac{1}{8}(\lambda+\delta\lambda_2)m_0^4f_d^2 -\frac{1}{4}m_0^4f_d+\frac{1}{8}(\lambda+\delta\lambda_2)H^4{f'}_d^2+\frac{1}{4}(\lambda+\delta\lambda_2)H^2m_0^2f_d{f'_d}-\frac{1}{2}H^2m_0^2{f'}_d\right)\right]
    \label{59eff}
\end{eqnarray}
The $v$-independent divergences appearing in the second line above can be absorbed in a cosmological constant counterterm in the gravitational action. The renormalised effective action, now regarded as the non-perturbative 1PI effective action, as it is no longer a function of the Green function, reads under the local or Hartree approximation 
\begin{eqnarray}
  &&  \Gamma_{\rm 1PI}[v]_{\rm Ren.}=-\int a^dd^dx\left[\frac12 (\nabla_{\mu} v)(\nabla^{\mu} v)
    +\frac{1}{2}m_0^2v^2+\frac{\lambda }{4!}v^4-\frac{\lambda}{8} f_{\rm fin}^2+\frac{1}{2}\int d{m}^2_{\rm dyn,eff}f_{\rm fin}\right]
    \label{59eff2}
\end{eqnarray}
where $f_{\rm fin}$ is given by \ref{y41a2}, \ref{y57'} and ${m}^2_{\rm dyn,eff}$ is given by \ref{y41a4}.  The effective potential corresponding to~\ref{59eff2} reads
\begin{eqnarray}
\bar{V}_{\rm eff}(\bar{v})= \frac{V_{\rm eff}(v)}{H^4}=\left(\frac{1}{2}\bar{m}_0^2\bar{v}^2+\frac{\lambda }{4!}\bar{v}^4-\frac{\lambda}{8} \bar{f}_{\rm fin}^2-\frac{1}{2}\int d\Bar{m}^2_{\rm dyn,eff}\bar{f}_{\rm fin}\right),
\label{59eff23}
\end{eqnarray}
where the bar over quantities denotes   scaling with respect to appropriate power of $H^2$.\\

\noindent
The explicit form of $\bar{V}_{\rm eff}(\bar{v})$ has to be found by numerical analyses. Note  from \ref{y41a4} that the dynamically generated mass also needs to be evaluated numerically in general. However, when $m^2_{\rm dyn, eff}$ is small compared to the Hubble rate, $m^2_{\rm dyn, eff}/H^2 \ll 1$, we may find out an analytic expression for it in the leading approximation,
\be
\bar{m}^2_{\rm dyn, eff}\simeq \frac{2\pi(\bar{m}_0^2+\lambda \bar{v}^2/2)+\sqrt{3\lambda+4\pi^2(\bar{m}_0^2+\lambda \bar{v}^2/2)}}{4\pi}
\label{DMs}
\ee
Note that for $\lambda =0$, we have $m^2_{\rm dyn, eff}= m_0^2$, i.e. there is no dynamically generated mass. This corresponds trivially to the fact that in the absence of interaction, there is no self energy. Next, for $m_0^2=0=\bar{v}$, we have $m^2_{\rm dyn, eff}= \sqrt{3\lambda}/4\pi$, showing that $m^2_{\rm dyn, eff}$ can {\it never} be vanishing. This reproduces the result of~\cite{Starobinsky:1994bd} and later confirmed by many others using different methods, e.g.~\cite{Bhattacharya:2023xvd} and references therein. Note also from \ref{y41a2} that for $m^2_{\rm dyn, eff}/H^2 \ll 1$, we have $f_{\rm fin}\sim m^{-2}_{\rm dyn, eff}$. Thus \ref{59eff23} shows that the effective potential remains bounded from below owing to this non-vanishing of $m^2_{\rm dyn, eff}$, even with $m_0^2=0$ and $v\to 0$.
\begin{figure}[h!]
\begin{center}
  \includegraphics[width=9.0cm]{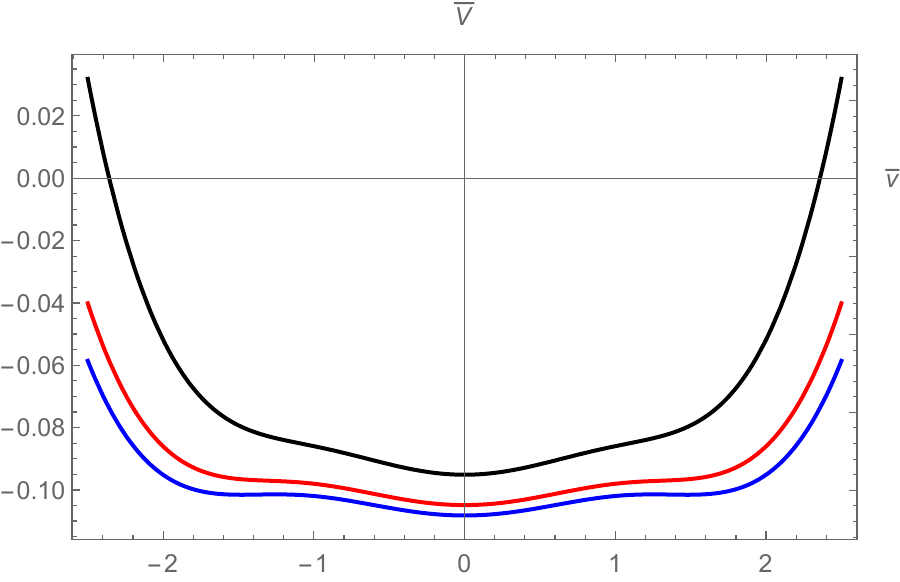}
 \caption{ \it \small  Variation of the effective potential \ref{59eff23}, with respect to the background field $v$, for negative rest mass squared $m_0^2 \sim -0.063H^2$. The bar over the quantities denote scaling with respect to appropriate power of the de Sitter Hubble rate, $H$.  The blue, red and black curves respectively correspond to $\lambda$ values $0.10$, $0.11$ and $0.15$ respectively. This non-trivial feature was reported first in~\cite{Arai:2012sh, Arai:2013jna, LopezNacir:2013alw}. }
  \label{Vplot2}
\end{center}
\end{figure}
\begin{figure}[h!]
\begin{center}
  \includegraphics[width=7.8cm]{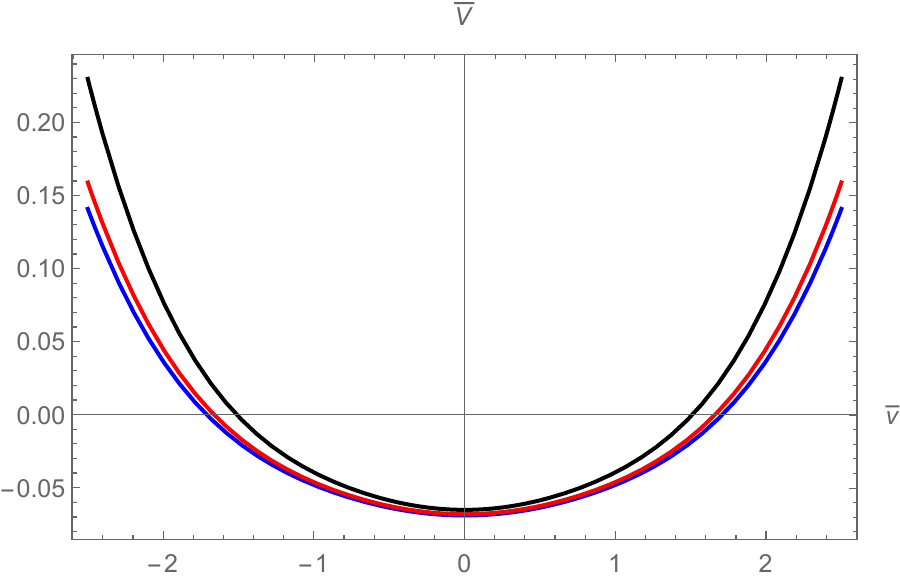}
  \includegraphics[width=7.5cm]{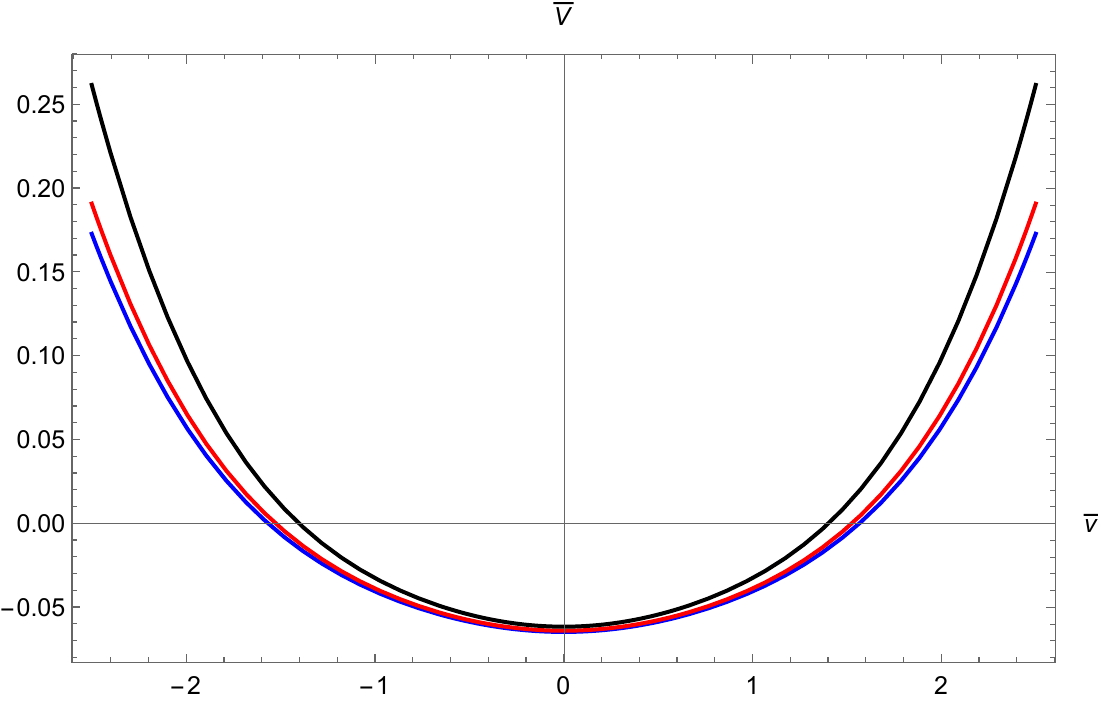}
 \caption{ \it \small Variation of the effective potential \ref{59eff23}, with respect to the background field $v$. The  The left and right set of curves correspond to $m^2_0=0$ and $m_0^2 \sim 0.01H^2$ respectively.  The blue, red and black curves respectively correspond to $\lambda$ values $0.10$, $0.11$ and $0.15$. }
  \label{Vplot1}
\end{center}
\end{figure}

\noindent
The variation of the effective potential, \ref{59eff23}, with respect to the background field $v$ has been depicted in \ref{Vplot2}, \ref{Vplot1}. In the first, we see a non-trivial behaviour for a negative rest mass squared, first reported in~\cite{Arai:2012sh, Arai:2013jna, LopezNacir:2013alw}.
\\

\noindent
Before we end this section, we note from our preceding discussion that in the presence of a background field, due to the $\lambda v^2/2$ term, the scalar cannot be treated as effectively massless, even if its rest mass is vanishing. However, given its special status in de Sitter pertaining to the isometry breaking and the non-existence of any smooth $m^2\to 0$ limit~\cite{Brunier:2004sb} (cf., the discussion in \ref{S2}), we wish to  briefly make some comments about the vacuum graphs of a massless and minimal scalar field in this background.  This corresponds to setting $v=0=m_0^2$. We wish to compute the 2PI vacuum graphs using the {\it tree level } or free propagator, \ref{props2}.

For the double bubble given by the first of \ref{f0}, we have using \ref{y6}
\begin{eqnarray}
i\Gamma_2\vert_{\rm 2-loop}=- \frac{\lambda}{2^3}\int a^d d^d x i\Delta^2_{++}(x,x)= -\frac{H^{4-2\e}\Gamma^2(2-\e)}{2^{7-2\e}\pi^{4-\e}\Gamma^2(1-\e/2)} \left(\frac{1}{\e^2}+\frac{2\ln a}{\e} + \ln^2 a \right)
\label{y43'}
\end{eqnarray}
The first divergence appearing above can be absorbed in a cosmological constant counterterm, whereas the second can be cancelled by the vacuum graph contribution generated by the one loop mass renormalisation counterterm, \ref{y7}.

\begin{figure}[h!]
\begin{center}
  \includegraphics[width=7.4cm]{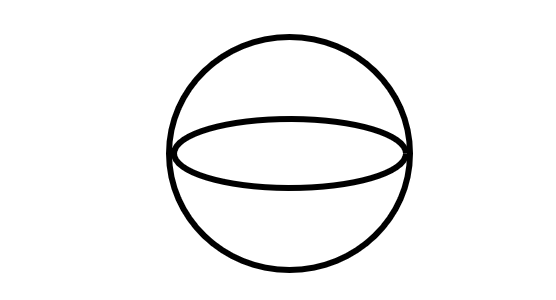}
 \caption{ \it \small Three loop (next to the leading order) 2PI vacuum diagram for $\phi^4$ self interaction. The other three loop diagram is a connected three-bubble, which is not 2PI. Although we have not considered  this graph for non-perturbative computations in this paper, we have computed its renormalised expression for a massless minimal scalar in \ref{A}, with the tree level propagtor,~\ref{props2}. See main text for discussion.}
  \label{fig3}
\end{center}
\end{figure}

The three loop vacuum graph is given by \ref{fig3}. Unlike the two loop case, it contains both local and non-local contributions.  The former gives divergences as well as subleading secular logarithms, whereas the latter yields the leading, deep infrared secular  contribution. Even though we are chiefly interested in the local parts of the vacuum graphs in this paper, we will compute the non-local part of \ref{fig3} as well. The corresponding renormalised contribution, computed in \ref{A}  using the Schwinger-Keldysh or in-in formalism, equals
\begin{eqnarray}
i\Gamma_2\vert_{\rm 3-loop, Ren.}= \frac{\lambda^2}{2^{10}\times 3\pi^6} \int a^4 d^4 x \left[ \frac16 \ln^4 a+ \ln^3a + {\cal O}(\ln^2 a)\right]
\label{y43''}
\end{eqnarray}
To the best of our knowledge, the diagram topology like \ref{fig3} has not been computed earlier in de Sitter.\\

\noindent
The issue of the secular logarithms does not bother us in \ref{59eff2}, simply because  we confined ourselves to two loop Hartree approximation there and second, the scalar field was not massless there due to the background field, no matter whether its rest mass is zero or not. We wish to show below that this will not be the case if we include the Yukawa interaction. Even though we shall focus on computing only the local part of $i\Gamma_2$, it will contain secular logarithm originating from the ultraviolet terms. Thus in order to reach any meaningful conclusion about the effective potential in this case, we must associate some finite value to such large logarithm.

\subsection{Inclusion of the Yukawa interaction}\label{S32}
%
\begin{figure}[h!]
\begin{center}
  \includegraphics[width=14.0cm]{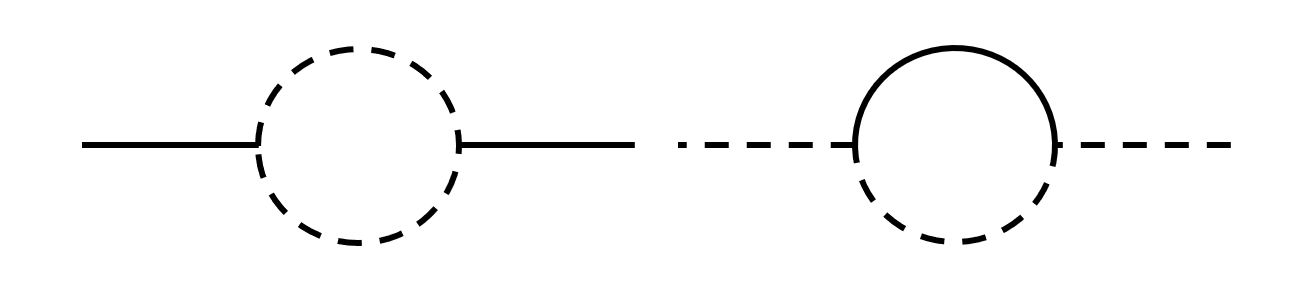}
 \caption{ \it \small One loop self energy diagrams for the Yukawa interaction. Solid and dashed lines respectively correspond to scalar and fermion propagators. The propagators are exact here. }
  \label{fig3s}
\end{center}
\end{figure}

\noindent
We now wish to add the effect of fermions to the effective potential of \ref{59eff23}.  The bare action is given by \ref{y4}. We assume that there is no background fermion field, i.e., the fermion is purely quantum. Then the general 2PI effective action is given by~\cite{Berges:2004yj},
\begin{eqnarray}
&&\Gamma_{\rm 2PI}[v, iG] = S[v]- \frac{i}{2} \ln {\rm det}\ i G (x,x) +  i\ln {\rm det }\ iS(x,x) +\frac{1}{2} {\rm Tr} \int (aa')^d d^dx d^d x' i \Delta^{-1}(x,x') iG(x',x)  \nonumber\\ && - {\rm Tr} \int (aa')^d d^dx d^d x' i S_0^{-1}(x,x') iS(x,x')    + i\Gamma_2 [v,iG] 
\label{y43}
\end{eqnarray}
where $i\Delta (x,x')$ and $iS_0(x,x')$ are respectively the tree level scalar and fermion propagators, whereas $iG(x,x')$ and $iS(x,x')$ are their exact forms.    The above expression explicitly reads for our theory
\begin{eqnarray}
&&\Gamma_{\rm 2PI}[v, iG] = -\int a^d d^d x \left[(1+\delta Z_1)\frac12 (\nabla_{\mu} v)(\nabla^{\mu} v) + \frac12 (m_0^2+\delta  m_{1}^2) v^2 +\frac{\delta \beta_1 v^3}{3!}+ \frac{\lambda_{1} v^4}{4!} + \delta \xi_{1} R v^2   \right]  \nonumber\\&& - \frac{i}{2} \ln {\rm det}\ i G (x,x)  +  i\ln {\rm det }\ iS(x,x) +\frac{1}{2} \int a^d d^dx \left[(1+\delta Z_2)\Box - (m_0^2+ \delta m^2_{2}) - 2\delta \xi_{2} R- \frac{\lambda_{2} v^2}{2} -\delta \beta_2 v \right]  iG(x,x)  \nonumber\\&&  - \int a^d d^d x \left[(1+\delta Z_f)i \slashed{\nabla} -M_0 -\delta M - g_2  v \right] iS(x,x)+ i\Gamma_2 [v,iG],
\label{y44}
\end{eqnarray}
where the 2PI vacuum contribution, \ref{f0}, in this case reads at two loop~\cite{Berges:2004yj},
\begin{eqnarray}
&&i\Gamma_{2}[v, iG] = - \frac{\lambda_{3}}{2^3} \int a^d d^d x \ i G^2(x,x) - \frac{i g_3^2}{2} {\rm Tr} \int (aa')^d d^dx d^d x'i S(x,x') iS(x',x) iG(x,x'),  
\label{y45}
\end{eqnarray}
and $g_2= g+\delta g_2$ and $g_3=g+\delta g_3$. \ref{fig3s} shows the one loop  scalar and fermion self energies corresponding to the two loop Yukawa vacuum graph. \\

\noindent
We begin with the equation of motion for the fermion propagator,
\begin{eqnarray}
&&  \left[ (1+\delta Z_f)i\slashed{\nabla} -M_0-\delta M - (g+\delta g_2)  v \right] iS(x,x') =  \frac{i\delta^d(x-x')}{a^d} - i(g+\delta g_3)^2 \int a''^d d^d x'' \left(i S(x,x'') iG(x,x'')\right) iS(x'',x') \nonumber\\
\label{y46}
\end{eqnarray}
As of the scalar field theory, we assume that the above equation is reduced to
\begin{eqnarray}
&&  \left[i \slashed{\nabla} -M_{\rm dyn, eff}  \right] iS(x,x') =  \frac{i\delta^d(x-x')}{a^d} + {\rm non-local ~terms}
\label{y47}
\end{eqnarray}
 where 
$$M_{\rm dyn, eff}= M_0+ g v + M_{\rm dyn}$$
The dynamical mass in~\ref{y47} originates from the integral  of \ref{y46} containing the fermion self energy. The non-local term in \ref{y47}, originates from the non-local part of the self energy, unlike the previously discussed case of scalar field theory in  the Hartree approximation. However, this non-local  contribution will not explicitly concern us in this work. Accordingly, we break the fermion propagator into two parts, and let $iS_l(x,x')$ be the part that satisfies the differential equation
$$\left[i \slashed{\nabla} -M_{\rm dyn, eff}  \right] iS_l(x,x') =  \frac{i\delta^d(x-x')}{a^d} $$
Thus $iS_l(x,x')$ is simply the propagator for a fermion with mass $M_{\rm dyn, eff}$. Now, for the local part of the self energy, we will  need the fermion propagator only at small scales, so that we obtain a $\delta$-function in the integrand in the self energy integral of \ref{y46}, shrinking the loop to a single point, and giving rise to the dynamically generated fermion rest mass.
We take the fermion to be light. Then the part of the propagator {\it relevant} for our present purpose  can be red off from~\ref{d7} of \ref{D} (\cite{Miao:2006pn}),
\begin{eqnarray}
&&  iS_l(x,x')= - \frac{i\Gamma(2-\e/2)}{2\pi^{2-\e/2}(aa')^{3/2-\e/2}} \frac{\slashed{\Delta x}}{(\Delta x^2)^{2-\e/2}} + \frac{\Gamma(1-\e/2)}{2^2\pi^{2-\e/2}(aa')^{3/2-\e/2}} \frac{a M_{\rm dyn,eff}}{(\Delta x^2)^{1-\e/2}}{\bf I}_{d\times d},
\label{y48}
\end{eqnarray}
where $\slashed{\Delta x}$ contains contraction with respect to the flat space gamma matrices. Note also that since we are working in local approximation, the propagators appearing above are all Feynman propagators, and we do not need to consider the Wightman functions necessary for the in-in formalism described in~\ref{A}, just like the $\phi^4$-theory in the two loop Hartree approximation discussed earlier. 

Using \ref{y48}, let us now compute the self energy term appearing in \ref{y46}. Using the expression \ref{d4'} for the scalar propagator and \ref{A4}, we have 
\begin{eqnarray}
&&  \int a''^d d^d x'' \left(i S(x,x'') iG(x,x'')\right) iS(x'',x')\nonumber\\ && =\int a^{-d} (aa'')^{3/2} d^d x'' \left[\frac{\mu^{-\e} \Gamma(1-\e/2) {\slashed
\partial} \delta^d (x-x'')}{2^4 \pi^{2-\e/2} (1-\e)\e} + \frac{i\mu^{-\e}\Gamma(1-\e/2)aM_{\rm dyn,eff}}{2^3\pi^{2-\e/2}(1-\e)\e}\delta^d(x-x'') \right]iS(x'',x') + {\rm non-local ~terms}\nonumber\\
\label{y49}
\end{eqnarray}
Note that  $iS_l(x,x')$, \ref{y48}, also yields non-local terms via the logarithms of \ref{A4}, which we are not considering here.
We now plug the above expression into \ref{y46}. The divergence associated with the first term on the right hand side of \ref{y49} can be tackled by a fermion field strength renormalisation counterterm, $\delta Z_f$. Following~\cite{Prokopec:2003qd}, we write
\begin{eqnarray}
&&  i\delta Z_f  \slashed{\nabla} iS(x,x')= i \delta Z_f \int d^d x'' a^{-d}(aa'')^{(d-1)/2}{\slashed \p}\delta^d (x-x'') iS(x'',x')
\label{y51}
\end{eqnarray}
where in $\slashed{\p}$, contraction with respect to the flat space gamma matrix has been used.
This immediately yields 
\begin{eqnarray}
&&   \delta Z_f = -\frac{\mu^{-\e}(g+\delta g_3)^2 \Gamma(1-\e/2)}{2^4\pi^{2-\e/2}(1-\e)\e}
\label{y52}
\end{eqnarray}
\ref{y46} can now be rewritten   after a little algebra as (suppressing the non-local terms for the sake of  brevity)
\begin{eqnarray}
&&  \left[i \slashed{\nabla}  -M_0 -\delta M -(g+\delta g_2)v - \frac{\mu^{-\e}(g+\delta g_3)^2\Gamma(1-\e/2) M_{\rm dyn, eff} }{2^3\pi^{2-\e/2} (1-\e) } \left( \frac{1}{\e}+\ln a\right) \right]iS_l(x,x')
=  \frac{i\delta^d(x-x')}{a^d} 
\label{y53}
\end{eqnarray}

We are yet to explicitly determine $\delta g_3$ and $M_{\rm dyn, eff}$ appearing in \ref{y52}, \ref{y53}. In order to do this, 
we group the non-derivative terms appearing on the left hand side of the above equation as 
\begin{eqnarray}
&& \delta M + \left( M_0 + gv + \frac{g^2 M_{\rm dyn, eff}}{2^3 \pi^2}\ln a \right) + \delta g_2 v  + \frac{\mu^{-\e}\Gamma(1-\e/2)M_{\rm dyn, eff}}{2^2\pi^{2-\e/2}(1-\e)}\left(g \delta g_3 \ln a + \frac{\delta g_3^2  \ln a }{2} + \frac{ (g+\delta g_3)^2}{2 \e} \right) 
\label{y53add1}
\end{eqnarray}
Thus we identify 
\begin{eqnarray}
&&M_{\rm dyn, eff} = M_0 + gv + \frac{g^2 M_{\rm dyn, eff}}{2^3 \pi^2}\ln a \nonumber\\ && \Rightarrow   M_{\rm dyn, eff} = \frac{M_0+g v}{1- \frac{g^2 \ln a}{2^3 \pi^2}}
\label{y53add2}
\end{eqnarray}
 Writing now $M_{\rm dyn, eff} = M_0 + gv+M_{\rm dyn}  $, we have the non-perturbative expression 
\begin{eqnarray}
M_{\rm dyn} =  \frac{g^2(M_0+g v)\ln a}{2^3\pi^2\left(1- \frac{g^2 \ln a}{2^3 \pi^2}\right)}
\label{y53add2'}
\end{eqnarray}
Thus for a massless fermion, there can be no dynamical generation of mass in de Sitter space if $v=0$, at least at two loop.  This is in contrast to a massless minimally coupled scalar and should be attributed to the conformal invariance of a massless fermion. The above expression also shows that in flat spacetime ($a=1$) there can be no dynamical mass generation.

We next substitute the expression for $M_{\rm dyn, eff}$ from \ref{y53add2} into the last term of \ref{y53add1}, to regroup them all as
\begin{eqnarray}
&&M_{\rm dyn, eff}+\left( \delta M + \frac{\mu^{-\e} M_0(g+\delta g_3)^2\Gamma(1-\e/2)}{2^3\pi^{2-\e/2}(1-\e)\e} \right) + \left( \delta g_2+\frac{\mu^{-\e}g(g+\delta g_3)^2\Gamma(1-\e/2)}{2^3\pi^{2-\e/2}(1-\e)\e}\right)v\nonumber\\&&
+  \frac{\mu^{-\e}\Gamma(1-\e/2)}{2^3\pi^{2-\e/2}(1-\e)} \left( (g+\delta g_3)^2 -g^2 +\frac{g^2 (g+\delta g_3)^2}{2^3\pi^{2}\e} \right) M_{\rm dyn, eff}\ln a 
\label{y53add3}
\end{eqnarray}
Since the above expression must equal $M_{\rm dyn, eff}$, the terms within each bracket must be vanishing. This yields the counterterms
\begin{eqnarray}
&& \delta g_3 =- g +\frac{g}{\left(1+ \frac{g^2 }{2^3 \pi^{2}\e} \right)^{1/2} }\nonumber\\
&& \delta g_2 = - \frac{\mu^{-\e}g(g+\delta g_3)^2\Gamma(1-\e/2)}{2^3\pi^{2-\e/2}(1-\e)\e} = -\frac{\mu^{-\e} g^3 \Gamma(1-\e/2)}{2^3 \pi^{2-\e/2}(1-\e)\e \left(1+ \frac{g^2 }{2^3 \pi^{2}\e} \right) }\nonumber\\
&& \delta M = -  \frac{\mu^{-\e}M_0(g+\delta g_3)^2\Gamma(1-\e/2)}{2^3\pi^{2-\e/2}(1-\e)\e} = -  \frac{\mu^{-\e} M_0g^2\Gamma(1-\e/2)}{2^3\pi^{2-\e/2}(1-\e)\e  \left(1+ \frac{g^2 }{2^3 \pi^{2}\e} \right)} 
\label{y50}
\end{eqnarray}
It is clear that the above counterterms are non-perturbative. Using the first of the above expression,  \ref{y52} becomes
\begin{eqnarray}
&&   \delta Z_f = -\frac{\mu^{-\e}g^2 \Gamma(1-\e/2)}{2^4\pi^{2-\e/2}\left(1+\frac{g^2}{2^3\pi^2\e}\right)(1-\e)\e}
\label{y52a}
\end{eqnarray}

\noindent
Before we proceed, let us  estimate the dynamically generated effective mass  of the fermion appearing in \ref{y53add2}. In an eternal de Sitter spacetime, clearly $M_{\rm dyn, eff}$ has many unsettling features, owing to the appearance of the $\ln a$ term. For example, it diverges when the denominator of \ref{y53add2} is vanishing, becomes negative afterward and then becomes asymptotically vanishing.  However, inflation cannot last forever, and we must put a cut off to the time scale of it and estimate the value of the logarithm at that time scale. For example, we can take it to be  just the standard number of $e$-foldings. Alternatively, and perhaps albeit naively, we may attempt to associate the $\ln a$ term 
with some ratio of the momentum/wavelength scale as follows.  Since we are dealing with the secular logarithm associated with the ultraviolet or local terms,  let us begin with a momentum $k$ at $a=1$. The proper momentum is $k/a$.  The ratio of the initial and late time values of this proper momentum  would be $k/(k/a)=(k a)/k=a$.  Inspired from this we write 
$$\ln a \sim \ln \frac{k_{\rm end}}{k_{\rm pivot}}$$
where $k_{\rm end} \sim 10^{23}{\rm Mpc}^{-1} $ is the scale of the horizon exit associated with the CMB, and $k_{\rm pivot}$ is some pivot scale which is taken as $\sim 0.05{\rm Mpc}^{-1}$, estimated from the reheating temperature, e.g.~\cite{Ebadi:2023xhq} and references therein.  We thus have $\ln a \sim 55.262$ at the end of the inflation. Note that this is also approximately the standard number of $e$-foldings.  We do {\it not} claim this estimation to be accurate. In particular as of now, we do not see in this framework any formal way to do resummation of this logarithm as opposed to e.g.~\cite{Kamenshchik:2020yyn, Bhattacharya:2023xvd}, where autonomous differential equations to sum perturbation series was constructed. On the other hand, the present formalism involves non-perturbative propagators only. Possibly  one then needs to think about additional equations to achieve any such formal resummation, if any. This remains as a possible caveat to our analysis. We also refer our reader to e.g~\cite{Bhattacharya:2022wjl} and references therein for discussion on the relationship of the deep infrared secular logarithms and ratio of different  momentum scale.

Putting things together now, we have at the end of inflation,
\be
M_{\rm dyn, eff} \simeq \frac{M_0+gv}{1- 0.7g^2}
\label{M-ln}
\ee 
Thus for example for the range $g \sim 0.1-0.5$, we see that the logarithm term contributes at most $0.7-17\%$ to $M_0+ gv$.  Due this reason we shall take $M_{\rm dyn, eff} \simeq M_0+gv$ in the following, and will ignore the dynamically generated fermion mass. In other words, the following discussions will be valid only for  small value of the Yukawa coupling.   A formal resummed result for  $M_{\rm dyn, eff}$ with generic $g$-values remains elusive to us so far.\\

\noindent
Let us now consider the equation of motion for the Green function of the scalar field 
\begin{eqnarray}
&&  \left[\Box - (m_0^2+\delta m^2_{2} )- 2\delta \xi_{2} R- \frac{(\lambda+\delta\lambda_{2}) v^2}{2}-\delta \beta_2 v\right] i G(x,x')=  \frac{i\delta^d (x-x')}{a^d} + \frac{(\lambda+\delta\lambda_3)}{2} iG(x,x) iG(x,x') \nonumber\\&&  + i(g+\delta g_3)^2 \int d^d x'' a''^d {\rm Tr }(iS(x,x'') iS(x'',x) )i  G(x'', x')\nonumber\\&& = \frac{i\delta^d (x-x')}{a^d}+\frac{(\lambda +\delta \lambda_3)}{2} \left[ m^2_{\rm dyn,eff} f_d + H^2 f'_d + f_{\rm fin}\right] iG(x,x')\nonumber\\&& + \frac{(g+\delta g_3)^2}{2\pi^2}\left[\frac{\mu^{-\e}  (M^2_{\rm dyn, eff}+H^2) \Gamma(1-\e/2)(1-\e/4)}{ \pi^{-\e/2}(1-\e)\e}  + (H^2+M^2_{\rm dyn, eff})\ln a\right]iG(x,x')+ \ {\rm non-local~terms} 
\label{y57}
\end{eqnarray}
where we have used \ref{y41a1}, \ref{y41a2}, \ref{B1}.  We will also take $M_{\rm dyn, eff}\simeq M_0+gv$, as argued above. We also have suppressed the scalar field strength renormalisation counterterm as it will not be necessary for our present purpose.

Before we proceed, first we note from \ref{y50} that the quantity $(g+\delta g_3)^2$ is not only finite, but also vanishingly small as ${\cal O}(\e)$. Thus the term ${(g+\delta g_3)^2/\e}$ is a non-vanishing and {\it finite} quantity as $\e \to 0$. 
This suggests that the finite quantities appearing within the square bracket  on the last line of \ref{y57} can safely be ignored, and the rest of the terms, containing ${(g+\delta g_3)^2/\e}$, are finite as $\e \to 0$. Note that analogous thing also happens for the pure quartic self interaction. For example, $\delta \lambda_2$ is finite and $(\lambda+\delta \lambda_2) \to 0$ as $\e \to 0$, \ref{y41a36}.  It can then be seen from \ref{58eff} that the product of this term and the divergent $(1+\lambda f_d/2)$ eventually yields ultraviolet finite non-vanishing terms.\\

\noindent
 It  seems {\it a priori} that we have two most natural ways to handle the issue of renormalisation of \ref{y57}. First,  we renormalise all the terms containing $(g+\delta g_3)^2/\e$ via mass, cubic and quartic coupling renormalisations (respectively, $\delta m_2^2$, $\delta \beta_2$, $\delta \lambda_2$ in \ref{y57}) through some {\it finite }counterterms. Alternatively, we keep them as it is in order to include them in the effective dynamical mass squared of the scalar field.  
 
However, following either of these paths leads to divergent term in the effective  action which cannot be renormalised away. This will be clear from our analysis below. In order to tackle this issue, we break the Yukawa coupling  term in \ref{y57} as
\be
(g+\delta g_3)^2  =(k+(1-k))(g+\delta g_3)^2
\label{break}
\ee
where $k$ is a number to be determined. We renormalise away all the terms containing $(1-k)(g+\delta g_3)^2$. Accordingly, the rest containing $k(g+\delta g_3)^2$ contributes to the scalar field's effective dynamical mass squared. With this, we  note from \ref{y57} the renormalisation conditions 
\begin{eqnarray}
&& \delta m_2^2 +\frac{(\lambda +\delta \lambda_3)}{2}\left[\left( {m}^2_0+ \frac{ k\mu^{-\e}(M_0^2+H^2) (g+\delta g_3)^2 \Gamma(1-\e/2)(1-\e/4)}{2\pi^{2-\e/2}(1-\e)\e}\right) f_d+ H^2 f'_d  \right]\nonumber\\&& + \frac{\mu^{-\e}(1-k) (g+\delta g_3)^2\Gamma(1-\e/2)(1-\e/4) (M_0^2+H^2)}{2\pi^{2-\e/2}\e(1-\e)}=0\nonumber\\
&& \delta \lambda_2 +(\lambda +\delta \lambda_3 )\left(\frac{\lambda }{2} + \frac{k \mu^{-\e} g^2 (g+\delta g_3)^2\Gamma(1-\e/2)(1-\e/4)}{2\pi^{2-\e/2}\e(1-\e)}\right)f_d+ \frac{\mu^{-\e}(1-k)g^2(g+\delta g_3)^2\Gamma(1-\e/2)(1-\e/4) }{\pi^{2-\e/2}\e(1-\e)}=0\nonumber\\&&
\delta \beta_2 + \frac{\mu^{-\e}(1-k)g(g+\delta g_3)^2\Gamma(1-\e/2)(1-\e/4)M_0 }{\pi^{2-\e/2}\e(1-\e)}+\frac{k\mu^{-\e}g M_0  (\lambda+\delta\lambda_3)(g+\delta g_3)^2f_d}{2\pi^{2-\e/2} (1-\e)\e}\Gamma(1-\e/2)(1-\e/4) =0,\nonumber\\ &&
\delta \lambda_3 = - \frac{\lambda^2 f_d}{2\left( 1+\frac{\lambda f_d}{2}\right)},\qquad \delta \xi_2=0. 
\label{y57a1}
\end{eqnarray}
  It is clear from the above expressions that either renormalising away the Yukawa contribution entirely or keeping it entirely corresponds respectively to $k=0$ or $k=1$ in the above equation. This would have been the simplest choice. However as we have mentioned above, doing so would lead to problem in cancelling divergence in the effective action, and as we will derive below, the {\it unique} choice is  $k=-1$. However until we prove it, let us keep $k$ generically.  Note that $k=-1$ partly takes away the Yukawa contribution, and the rest remains as finite contribution. As a physical consequence of the above renormalisation, the   dynamical effective scalar mass squared acquires an Yukawa contribution given by, 
\be
m^2_{\rm dyn, eff}= m_0^2+\frac{\lambda v^2 }{2} +\frac{\lambda f_{\rm fin}}{2} + \frac{\mu^{-\e}k (g+\delta
g_3)^2((M_0+gv)^2+H^2)\Gamma(1-\e/2)(1-\e/4)}{2\pi^{2-\e/2}(1-\e)\e}
\label{y57a4}
\ee
 Recall that by the virtue of \ref{y50}, $(g+\delta g_3)^2/\e$ is a finite quantity. Note also the appearance of the Fermion mass squared, $(M_0+g v)^2$ in the above expression. 

\begin{figure}[h!]
\begin{center}
  \includegraphics[width=10.0cm]{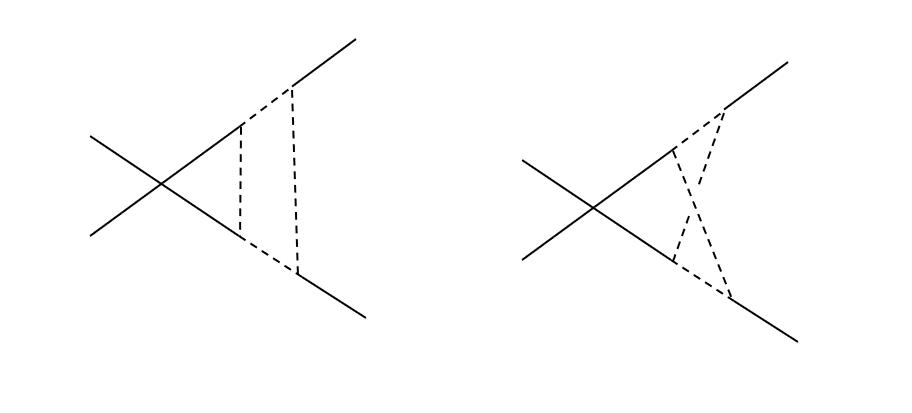}
 \caption{ \it \small A non-trivial scalar quartic vertex function correction  and its corresponding resummed counterterm generated by the 2PI formalism, \ref{y57a2}. Solid and dashed lines represent respectively, scalar and fermions.}
  \label{fig3'}
\end{center}
\end{figure}

Using \ref{y41a2}, \ref{y50},  the counterterms appearing in \ref{y57a1} can explicitly be written as
\begin{eqnarray}
&& \delta m_2^2 = \frac{\lambda H^{-\e}}{2^4\pi^{2-\e/2}\e \left( 1-\frac{\lambda H^{-\e}}{2^4\pi^{2-\e/2}\e} \right)}\left( m_0^2+H^2 (2-\gamma \e )   +\frac{ k \mu^{-\e}(M_0^2+H^2) g^2 \Gamma(1-\e/2)(1-\e/4)}{2\pi^{2-\e/2}(1-\e)\e\left(1 +\frac{g^2}{2^3\pi^2 \e}\right)}\right)\nonumber\\ &&   - \frac{\mu^{-\e}(1-k)g^2(M_0^2+H^2)\Gamma(1-\e/2)(1-\e/4)}{2\pi^{2-\e/2}\e(1-\e)\left( 1+\frac{g^2}{2^3\pi^{2}\e} \right)},\nonumber\\&&
  \delta \beta_2 = -\frac{\mu^{-\e} (1-k) g^3M_0\Gamma(1-\e/2)(1-\e/4)}{\pi^{2-\e/2}\e(1-\e)\left( 1+ \frac{g^2}{2^3\pi^2\e}\right)}+  \frac{k (\mu H)^{-\e} M_0 \lambda g^3 \Gamma(1-\e/2)(1-\e/4)}{2^{4-\e}\pi^{4-\e}(1-\e)\e^2\left(1+\frac{g^2}{2^3\pi^2\e} \right)\left( 1-\frac{\lambda H^{-\e}}{2^{4-\e}\pi^{2-\e/2}\e}\right)},    \nonumber\\&&
\delta \lambda_2= \frac{\lambda H^{-\e}}{2^{4-\e}\pi^{2-\e/2}\e\left( 1-\frac{\lambda H^{-\e}}{2^{4-\e}\pi^{2-\e/2}\e}  \right)} \left(\lambda+ \frac{k\mu^{-\e} g^4 \Gamma(1-\e/2)(1-\e/4)}{\pi^{2-\e/2}\e(1-\e)\left(1+\frac{g^2}{2^3\pi^2\e} \right)} \right)- \frac{\mu^{-\e} (1-k)g^4\Gamma(1-\e/2)(1-\e/4)}{\pi^{2-\e/2}\e(1-\e)\left( 1+\frac{g^2}{2^3\pi^{2}\e} \right)}, \nonumber\\ &&
\delta \lambda_3 = \frac{\lambda^2 H^{-\e}}{2^{4-\e}\pi^{2-\e/2}\e\left( 1-\frac{\lambda H^{-\e}}{2^{4-\e}\pi^{2-\e/2}\e}  \right)},
\qquad \qquad \qquad \delta \xi_2=0.
\label{y57a2}
\end{eqnarray}
Note that $\delta \lambda_2\neq \delta \lambda_3$ here, unlike the earlier case of pure scalar field theory. The structure of the counterterms appearing above suggest various non-trivial self energy and vertex corrections and their resummation. For example, the last term on the right hand side in the expression for $\delta \lambda_2$ corresponds to the renormalisation of the box diagram corresponding to the scalar four point function in the Yukawa theory. On the other hand, the second term on the right hand side of the same corresponds to the resummed renormalisation of the quartic scalar vertex correction due to the insertion of the Yukawa vertices, \ref{fig3'}.  In particular, note from \ref{y57a2} that the value of $k$ can decide the diagrams generated. For example for $\delta \lambda_2$, $k=1$ omits the Yukawa box diagram totally, whereas $k=0$ omits the diagrams of \ref{fig3'}. $k=-1$ on the other hand, keeps all the diagrams corresponding to  all the counterterms.

\noindent
With this, \ref{y57} can be rewritten as (suppressing the non-local terms)
\begin{eqnarray}
(\Box -m^2_{\rm dyn, eff})iG_{\rm loc}(x,x') = \frac{i\delta^d (x-x')}{a^d},
\label{y57a3}
\end{eqnarray}
with $m^2_{\rm dyn, eff}$ being given by \ref{y57a4}.

\bigskip

\noindent
We are yet to clarify the necessity of introducing the constant $k$ and its value appearing in 
\ref{y57a1}, \ref{y57a4}, \ref{y57a2}. In order to see this, let us now consider the effective action \ref{y43}, which, after using the equations of motion for the propagators,  \ref{y46}, \ref{y57}, becomes
\begin{eqnarray}
&&\Gamma_{\rm 2PI}[v, iG] = -\int a^d d^d x \left[\frac12 (\nabla_{\mu} v)(\nabla^{\mu} v) + \frac12 (m_0+\delta  m_{1}^2 )v^2 + \frac{(\lambda +\delta \lambda_{1}) v^4}{4!} + \frac{\delta \beta_1 v^3}{3!}+ \delta \xi_{1} R v^2 + \gamma_{\rm tad}v\right] \nonumber\\&&- \frac12 \int d^d x a^d \int d m^2_{\rm dyn, eff}\ iG(x,x) + \int d^d x a^d \int d M_{\rm dyn, eff}\ iS(x,x) \nonumber\\&& + \frac{\lambda+\delta \lambda_3}{2^3}\int a^d d^d x \ i G^2(x,x) + i (g+\delta g_3)^2 {\rm Tr} \int (aa')^d d^dx d^d x'i S(x,x') iS(x',x) iG(x,x')
\label{y64}
\end{eqnarray}
We now wish to compute the vacuum loop integrals appearing in the last line of \ref{y64}. Let us for the sake of brevity, abbreviate in \ref{y57a4},
\be
\frac{\mu^{-\e} (g+\delta g_3)^2\Gamma(1-\e/2)(1-\e/4)}{2\pi^{2-\e/2}(1-\e)\e}= C \ ({\rm say})
\label{C-def}
\ee
Using next \ref{y41a1}, \ref{y57a4} and \ref{B1} of \ref{B}, we obtain for \ref{y64} after a little algebra
\begin{eqnarray}
&&\frac{\lambda+\delta \lambda_3}{2^3}\int a^d d^d x \ i G^2(x,x) + i (g+\delta g_3)^2 {\rm Tr} \int (aa')^d d^dx d^d x'i S(x,x') iS(x',x) iG(x,x') \nonumber\\ && = \frac{\lambda}{2^3(1+\lambda f_d/2)}\int a^d d^d x \left[ \left( m_0^2f_d + H^2 f'_d \right)^2 + \frac{\lambda^2 v^4}{4}f_d^2 +k^2C^2\left((M_0+gv)^2+H^2\right)^2f_d^2 + \left(1+\frac{\lambda f_d}{2}\right)^2f_{\rm fin}^2\right. \nonumber\\&& \left. + \lambda f_d (m_0^2f_d+H^2 f'_d)v^2 + 2kCf_d (m_0^2f_d+H^2 f'_d)\left((M_0+gv)^2+H^2\right)+2(m_0^2f_d+H^2 f'_d)\left(1+\frac{\lambda f_d}{2}\right)f_{\rm fin}\right. \nonumber\\ &&\left. + \lambda  f_d \left(1+\frac{\lambda f_d}{2}\right)v^2 f_{\rm fin} + \lambda v^2 kC f_d^2 \left((M_0+gv)^2+H^2\right) + 2kC \left((M_0+gv)^2+H^2\right)\left(1+\frac{\lambda f_d}{2}\right)f_{\rm fin} f_d\right]\nonumber\\&&
+\frac{C}{2}\int a^d d^d x
\left(H^2+(M_0+gv)^2\right)\left( 1+ \e\ln a+ \frac12\e^2\ln^2 a+{\cal O}(\e^3) \right)\left[\left(m_0^2+\frac{\lambda v^2}{2}+\frac{\lambda f_{\rm fin}}{2} +Ck\left(H^2+(M_0+gv)^2\right)  \right)f_d\right. \nonumber\\ && \left.+ H^2 f'_d + f_{\rm fin}\right] 
\label{y65}
\end{eqnarray}
 In the above expression the most problematic divergence appears in the penultimate line, given by
$$ \frac{C\lambda}{4}\left( H^2 + (M_0+gv)^2\right)f_{\rm fin} f_d,$$
Given the structure of $f_{\rm fin}$, \ref{y41a1}, the above divergence cannot be absorbed by any standard counterterm we know of. This is an overlapping divergence and is absent in the traditional  computations.  Recall also that we do not have the freedom here to add any further graph in the effective  action to attempt a cancellation.  Putting things together, we now make the choice 
$$k=-1,$$
 in \ref{y57a1}, \ref{y57a2} and \ref{y65}, so that the above overlapping divergence gets cancelled by the last term of the fourth line of \ref{y65}. It is the necessity of cancellation of this divergence that led us to introduce the constant $k$ earlier. {\it Any}  choice other than $k=-1$ (like $k=1$ or $k=0$) would not have served our purpose. Note also that this problem would not have arisen if the sign in front of the last term on the right hand side of  the effective action \ref{y64} was opposite. However, we have checked that there is no error in any sign. The most important consequence of this renormalisation is the particular form of the scalar dynamical mass, \ref{y57a4}, which will contribute to the non-trivial characteristics of the effective potential as follows.  \\

\noindent
Putting things together, we now rewrite \ref{y65} as
\begin{eqnarray}
&&\frac{\lambda+\delta \lambda_3}{2^3}\int a^d d^d x \ i G^2(x,x) + i (g+\delta g_3)^2 {\rm Tr} \int (aa')^d d^dx d^d x'i S(x,x') iS(x',x) iG(x,x') \nonumber\\ && =  \frac{\lambda}{2^3(1+\lambda f_d/2)}\int a^d d^d x \left[ \left( m_0^2f_d + H^2 f'_d \right)^2 + \frac{\lambda^2 v^4}{4}f_d^2 +C^2\left((M_0+gv)^2+H^2\right)^2f_d^2 + \left(1+\frac{\lambda f_d}{2}\right)^2f_{\rm fin}^2\right. \nonumber\\&& \left. + \lambda f_d (m_0^2f_d+H^2 f'_d)v^2 - 2Cf_d (m_0^2f_d+H^2 f'_d)\left((M_0+gv)^2+H^2\right)+2(m_0^2f_d+H^2 f'_d)\left(1+\frac{\lambda f_d}{2}\right)f_{\rm fin}\right. \nonumber\\ &&\left. + \lambda  f_d \left(1+\frac{\lambda f_d}{2}\right)v^2 f_{\rm fin} - \lambda v^2 C f_d^2 \left((M_0+gv)^2+H^2\right) \right]\nonumber\\&&
+\frac{C}{2}\int a^d d^d x
\left[H^2+(M_0+gv)^2\right]\left[\left(m_0^2+\frac{\lambda v^2}{2}-C\left(H^2+(M_0+gv)^2\right)  \right)f_d+ H^2 f'_d + f_{\rm fin}\right. \nonumber\\ &&\left.
- \left( m_0^2+\frac{\lambda v^2}{2}+\frac{\lambda f_{\rm fin} }{2}-C\left(H^2+(M_0+gv)^2\right)+2H^2 \right)\frac{\ln a}{2^3\pi^2}\right] 
\label{y65a}
\end{eqnarray}

\noindent
 We next compute for \ref{y64} using the unrenormalised coincidence limit expressions \ref{y41a1} and \ref{d7},
\begin{eqnarray}
&& \frac12 \int d^d x a^d \int d m^2_{\rm dyn, eff}\ iG(x,x) -\int d^d x a^d \int d M_{\rm dyn, eff}\ iS(x,x)\nonumber\\&&
= \int d^d x a^d \left[ \frac{m^4_{\rm dyn, eff}}{4}f_d+ \frac{H^2 m^2_{\rm dyn, eff}}{2} f'_d + \frac12  \int d m^2_{\rm dyn, eff} f_{\rm fin} \right]  \nonumber\\&&
 -\int d^d x a^d \left[ \left( -\frac{2}{\e}+\frac32 +\gamma \right) \frac{M^2_{\rm dyn,eff} H^{2-\e}}{2^{1-\e/2}\pi^{2-\e/2}} +   \left( -\frac{2}{\e}-\frac32 +\gamma \right) \frac{M^4_{\rm dyn,eff} H^{-\e}}{2^{2-\e/2}\pi^{2-\e/2}} + \int dM_{\rm dyn, eff} F_{\rm fin}   \right],    
\label{y66}
\end{eqnarray}
where we have abbreviated from \ref{d7},
\begin{eqnarray}
F_{\rm fin}= \frac{M_{\rm dyn, eff}H^2}{\pi^2} \left( 1+ \frac{M_{\rm dyn, eff}^2}{H^2}\right)\left[\psi\left(1+ \frac{iM_{\rm dyn, eff}}{H}\right)+ \psi\left(1- \frac{iM_{\rm dyn, eff}}{H} \right)\right], 
\label{y67}
\end{eqnarray}
where $\psi$ is the digamma function. Also recall that we have taken $M_{\rm dyn, eff} \simeq  M_0 +gv$.\\

\noindent
In order to find out the effective action \ref{y64}, we next subtract \ref{y66} from \ref{y65a}, and  obtain after a little algebra
\begin{eqnarray}
&&\int a^4 d^4 x\left[\int dM_{\rm dyn, eff} F_{\rm fin}  -\frac12  \int d m^2_{\rm dyn, eff} f_{\rm fin}+ \frac{\lambda f^2_{\rm fin}}{8} \right.\nonumber\\ && \left. +\frac{C}{2}\left[H^2+(M_0+gv)^2\right]\left\{f_{\rm fin}-\left( m_0^2+\frac{\lambda v^2}{2}+\frac{\lambda f_{\rm fin} }{2}-C\left(H^2+(M_0+gv)^2\right)+2H^2 \right)\frac{\ln a}{2^3\pi^2}\right\} \right]  \nonumber\\ && + \int a^d d^d x\left[\left\{\frac{\lambda ((m_0^2 f_d + H^2 f'_d) - C(M_0^2+H^2)f_d)^2}{2^3(1+\lambda f_d/2)}-\frac{m_0^2}{2}\left(m_0^2f_d+2H^2f'_d \right)+ \frac{M_0^2H^{-\e}}{2^{1-\e/2}\pi^{2-\e/2}}\left(\frac{M_0^2}{2}\left(-\frac{2}{\e}-\frac32+\gamma \right)\right. \right.\right.\nonumber\\ && \left.\left. \left.+ H^2\left(-\frac{2}{\e}+\frac32+\gamma \right)\right) +\frac{C}{2}(H^2+M_0^2)\left((m_0^2-C(H^2+M_0^2))f_d+H^2 f'_d\right) \right\} \right.\nonumber\\ &&\left. - \frac{\lambda(m_0^2 f_d+H^2 f'_d)v^2}{4(1+\lambda f_d/2)} - \frac{\lambda^2 f_d v^4}{16(1+\lambda f_d/2)} \right] \nonumber\\&&
+\int a^d d^d x M_0 g \left[\frac{H^{-\e}}{2^{-\e/2}\pi^{2-\e/2}}\left\{H^2\left(-\frac{2}{\e}+\frac32+\gamma \right)+M_0^2\left(-\frac{2}{\e}-\frac32+\gamma \right) \right\} \right. \nonumber\\ && \left. +\frac{\lambda C}{2(1+\lambda f_d/2)}\left(C(H^2+M_0^2)-(m_0^2f_d+H^2 f'_d)f_d\right)+C \left( (m_0^2f_d+H^2 f'_d)-2C(H^2+M_0^2)f_d \right) \right] v\nonumber\\&&
+\int d^d x a^d  \left[ \frac{g^2H^{-\e}}{2^{1-\e/2}\pi^{2-\e/2}}\left\{H^2\left(-\frac{2}{\e}+\frac32+\gamma \right)+3M_0^2\left(-\frac{2}{\e}-\frac32+\gamma \right) \right\}\right. \nonumber\\ && \left. +\frac{\lambda}{4(1+\lambda f_d/2)} \left((3M_0^2+H^2)C^2 g^2 f_d^2-Cf_d g^2(m_0^2 f_d+H^2 f'_d)- \frac{\lambda C f_d^2}{2}(M^2+H_0^2)\right) \right. \nonumber\\ && \left. +\frac{C}{2}\left(g^2\left( (m_0^2-C(H^2+M_0^2)) f_d+H^2 f'_d\right) + \frac{\lambda}{2}(H^2+M_0^2)f_d-4g^2CM_0^2f_d\right)\right]v^2\nonumber\\ &&
+ \int a^d d^d x  M_0g\left[Cf_d\left\{ -\frac{\lambda^2 f_d}{4(1+\lambda f_d/2)}+\left(\frac{\lambda}{2}-2Cg^2 \right)\right\}+ \left(-\frac{2}{\e}-\frac32+\gamma\right)\frac{ g^2 H^{-\e}}{2^{-\e/2}\pi^{2-\e/2}} \right]v^3\nonumber\\&&
+ \int a^d d^d x g^2 \left[ \frac{\lambda f_d^2 C}{1+\lambda f_d/2}(Cg^2 -\lambda)+ \frac{C}{2}\left(\frac{\lambda}{2}-Cg^2 \right)f_d+ \left(-\frac{2}{\e}-\frac32+\gamma\right)\frac{ g^2 H^{-\e}}{2^{2-\e/2}\pi^{2-\e/2}}\right]v^4, 
\label{y67a}
\end{eqnarray}
where $M_{\rm dyn,eff}$ and $F_{\rm fin}$ is given by \ref{y67}, $m^2_{\rm dyn, eff}$ by \ref{y57a4} (with $k=-1$), $f_{\rm fin}$, $f_d$ and $f'_d$ by \ref{y41a2} and \ref{y57'} and $C$ is defined in \ref{C-def}. 
The first two lines of the above equation are finite. The fifth line gives the divergences corresponding to the quartic self interaction for mass and coupling constant renormalisation. The remaining lines give the divergences corresponding to the Yukawa  or overlapping Yukawa-$\phi^4$ interactions, and they all vanish in the absence of the Yukawa coupling. When \ref{y67a} is plugged into \ref{y64}, the terms proportional to $v$, $v^2$, $v^3$ and $v^4$ respectively give the tadpole ($\gamma_{\rm tad}$), the mass renormalisation ($\delta m_1^2$), the cubic coupling ($\delta \beta_1 $) and the quartic coupling ($\delta \lambda_1$) counterterms.  We  set $\delta \xi_1=0$. Note also that for vanishing fermion rest mass, we have $\delta \beta_1=0= \gamma_{\rm tad}$. Finally, the third and fourth lines are field independent and hence must be absorbed in a cosmological constant counterterm in the gravitational action. \\

\noindent
Putting things together, we now have the renormalised 1PI effective action, only  as a function of the background scalar field,
\begin{eqnarray}
&&\Gamma_{\rm 1PI}[v]_{\rm Ren} = -\int a^d d^d x \left[\frac12 (\nabla_{\mu} v)(\nabla^{\mu} v) + \frac12 m_0v^2 + \frac{\lambda  v^4}{4!}- \int dM_{\rm dyn, eff} F_{\rm fin} (M_{\rm dyn, eff}) +\frac12  \int d m^2_{\rm dyn, eff} f_{\rm fin}(m^2_{\rm dyn, eff} )\right.\nonumber\\ && \left. - \frac{\lambda f^2_{\rm fin}}{8} -\frac{C}{2}\left[H^2+(M_0+gv)^2\right]\left\{f_{\rm fin}-\left( m_0^2+\frac{\lambda v^2}{2}+\frac{\lambda f_{\rm fin} }{2}-C\left(H^2+(M_0+gv)^2\right)+2H^2 \right)\frac{\ln a}{2^3\pi^2}\right\} \right]\nonumber\\ && + \ {\rm non-local~ contributions.}
\label{y71}
\end{eqnarray}
The first line of the above equation corresponds to the standard tree plus one loop effective action, whereas the second line gives the contribution due to integrating out the 2PI vacuum graphs, \ref{f0}.
In the absence of the fermion, the above reduces to the scalar field theory result, \ref{59eff2}, whereas setting  $a=1$ yields the result of the flat spacetime.   \\

What will be the differences of the renormalisation procedure and the final result if we instead  compute the two loop effective action for the $\phi^4$-Yukawa theory via standard 1PI perturbation theory? This issue has been outlined very briefly in \ref{E}. Chiefly, owing to the fact that one uses the tree level propagators in such computations, we show that the renormalisation procedure is qualitatively very different in that case. In particular,  we show that there is no necessity of the constant $k$, as of \ref{y57a1}. Second, we show that the finite part of the two loop effective action is also different, and in fact the secular logarithm is quadratic there opposed to  \ref{y71}. \\

\noindent
Now, even though renormalised, \ref{y71} has the obvious problematic feature associated with the secular logarithm, which grows monotonically, and eventually becomes large at late times. As earlier, we will assign with it the value we estimated, $\ln a \sim 55.262 $ (cf, the discussion below \ref{y52a}). In particular unlike what we did for  fermion's effective dynamical mass,  since this logarithm  is not multiplied with any $g^2$, we cannot ignore it compared to the $f_{\rm fin}$ term appearing in the curly bracket of \ref{y71}.   Substituting this {\it estimated} value of $\ln a$ into \ref{y71}, we have
\begin{eqnarray}
\Gamma_{\rm 1PI}[v]_{\rm Ren} =&& -\int a^d d^d x \left[\frac12 (\nabla_{\mu} v)(\nabla^{\mu} v) + \frac12 m_0v^2 + \frac{\lambda  v^4}{4!}- \int dM_{\rm dyn, eff} F_{\rm fin}  +\frac12  \int d m^2_{\rm dyn, eff} f_{\rm fin}\right.\nonumber\\ && \left. - \frac{\lambda f^2_{\rm fin}}{8} -\frac{C}{2}\left[H^2+(M_0+gv)^2\right]\left\{f_{\rm fin}-0.7\left(\frac{\lambda v^2}{2}+\frac{\lambda f_{\rm fin} }{2}-C\left(2M_0gv+g^2v^2\right)\right)\right\} \right] \nonumber\\ &&+ \ {\rm non-local~ contributions},
\label{y72}
\end{eqnarray}
where we have absorbed further some field independent constant terms into redefinition of the cosmological constant in the gravitational action. \ref{y72} is the final result of this paper. Note that owing to our estimation of the secular logarithm, the above effective action is valid only towards the end of the inflation. At some earlier stages, one may just ignore the $\ln a$ term. We will briefly come back to the issue of $\ln a$ towards the end of this Section, and will argue that use of a  recently proposed renormalisation group inspired autonomous equation to assign a non-perturbative value to this secular logarithm yields qualitatively similar result.   \\

\noindent
Let us now specialise to the zero rest mass cases, $M_0=0=m_0$, so that the local effective potential corresponding to \ref{y72} reads
\begin{eqnarray}
V_{\rm eff, loc}= && \frac{\lambda  v^4}{4!}- \int dM_{\rm dyn, eff} F_{\rm fin}  +\frac12  \int d m^2_{\rm dyn, eff} f_{\rm fin} - \frac{\lambda f^2_{\rm fin}}{8}+ \frac{0.7 C}{4}\left(\lambda-2Cg^2 \right)v^2\left(H^2+g^2v^2\right)\nonumber\\&&-\frac{C}{2}f_{\rm fin}\left( 1-\frac{0.7\lambda}{2}\right)\left(H^2+g^2v^2\right),
\label{y73}
\end{eqnarray}
where $C$ is given by \ref{C-def}, and we recall once again that by the virtue of the first of \ref{y50}, it is absolutely finite as $\e \to 0$. By setting  $g=0$, we reproduce the result for the pure scalar field theory, \ref{59eff23}.\\

\begin{figure}[h!]
\begin{center}
  \includegraphics[width=6.0cm]{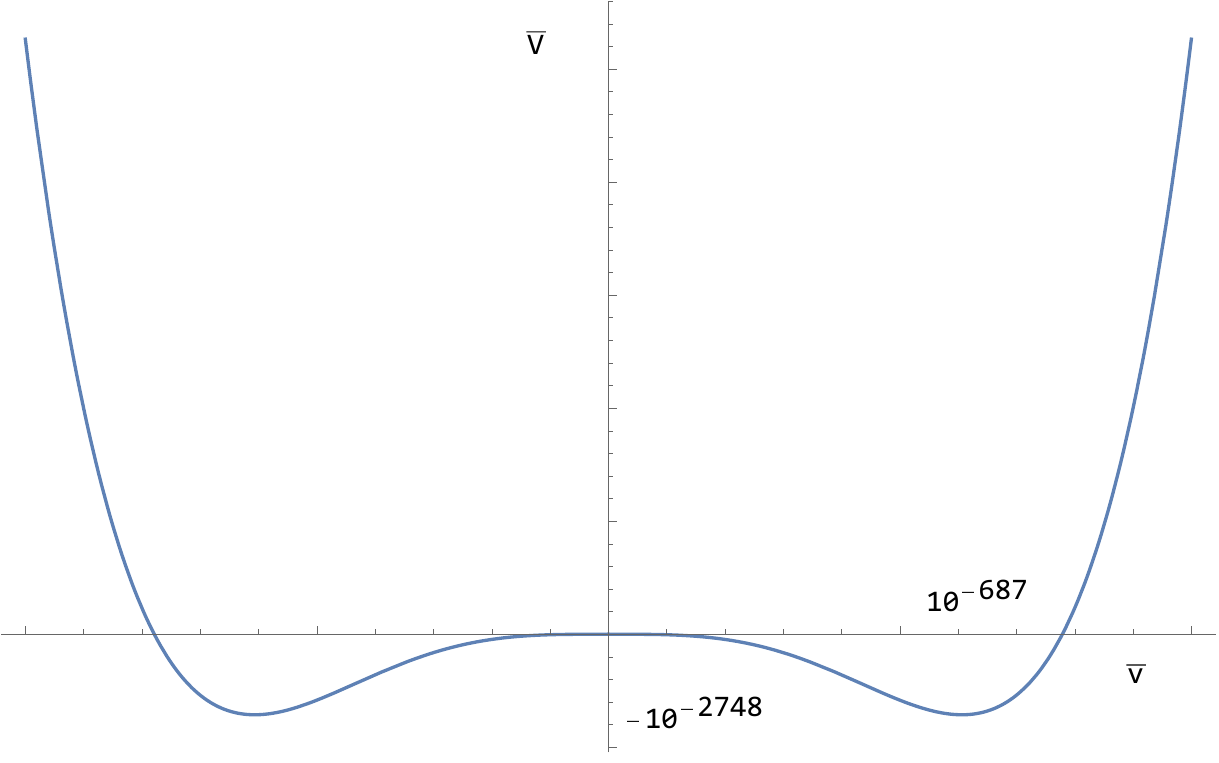}
 \caption{ \it \small Spontaneous symmetry breaking feature of  the flat spacetime  limit ($H\to 0$) of the  effective potential, \ref{y74}, with respect to the background scalar field $v$. Bar over the quantities denotes that they are dimensionless.  See main text for discussion. We have taken the coupling values $g=0.06$ and $\lambda=0.1$.}
  \label{Vplot133}
\end{center}
\end{figure}

\noindent
Before we investigate the behaviour of \ref{y73}, let us first take the high field limit of \ref{y73}, $|v|/H \gg 1$. Using the asymptotic expansion for the digamma function,  
$$\psi (1+x)\vert_{|x|\gg 1} \simeq \ln x + {\cal O}\left(\frac{1}{x}\right),$$
we have from \ref{y67}, \ref{y57a4} (with $k=-1$) the leading expression 
\begin{eqnarray}
&& V_{\rm eff, loc}\vert_{|v|/H\gg1} \simeq \left[ \frac{\lambda  v^4}{4!} + \frac{1}{2\pi^2}\left(\frac{(\lambda -8g^2)^2}{64}-g^4 \right) v^4 \ln \frac{|v|}{H}\right]
\label{y74}
\end{eqnarray}
Thus if we want our potential to be bounded from below in order to avoid any runaway disaster for the system, we must have,
\be
\lambda \gtrsim 16 g^2
\label{y75}
\ee
 Although this high field limit~\ref{y74} is  similar to that of the standard one loop effective action,  the coefficients of the $v^4\ln v$ term contains  contribution beyond one loop. Note  from \ref{y75} that if we want $\lambda$ to be less than unity, we must have $g \lesssim 0.3$. This is consistent with our argument $M_{\rm dyn, eff}\simeq M_0+gv$, we made earlier for the fermion (cf, the discussion below \ref{y52a}).  Finally, we also note that $|v|/H \gg1 $  can also be interpreted as the flat spacetime $(H\to 0)$ limit of \ref{y73}. The constraint \ref{y75} ensures that \ref{y74} has the standard Coleman-Weinberg spontaneous symmetry breaking feature, as depicted in \ref{Vplot133}.  Also note that since the violation of the upper bound of \ref{y75} makes $V_{\rm eff, loc}$ unbounded from below, it may lead to vacuum decay via tunnelling.     However, this scenario is physically unacceptable, for no late time equilibrium state of the scalar field can be reached then.  The effective potential will become negative and huge,  creating a large backreaction  to eventually push the universe towards a runaway disaster in the anti-de Sitter phase.

We also note here that for a zero rest mass scalar, the  small field limit of the corresponding one loop effective action diverges as $v\to 0$, in the standard perturbative approach. The  one loop effective action corresponds to the integral over the mass terms in \ref{y71}. In the perturbative case, we do not have the dynamically generated mass, and have instead simply, $m^2_{\rm dyn, eff}=\lambda v^2/2$. Using then \ref{y41a2}, it is easy to see that the perturbative one loop effective action diverges  like $\sim \ln |v|$, as $v \to 0$.  This divergence corresponds to the fact that setting $v=0$ for the present case makes the scalar {\it purely} massless. Thus as $v \to 0$, we are basically taking the massless limit of a massive scalar, which does not exist in de Sitter. In other words,  we must not ignore the scalar's dynamically and non-perturbatively generated rest mass.  Indeed, as we discussed below \ref{DMs}, $m^2_{\rm dyn, eff}$ can {\it never} be vanishing by the virtue of the dynamical mass, eventually making the effective potential divergence free for $m_0^2=0$ and $v\to 0$. Alternatively, in the absence of a background field, we must work with the propagator appropriate for the purely massless minimally coupled scalar, \ref{props2}.   \\

\begin{figure}[h!]
\begin{center}
  \includegraphics[width=7.8cm]{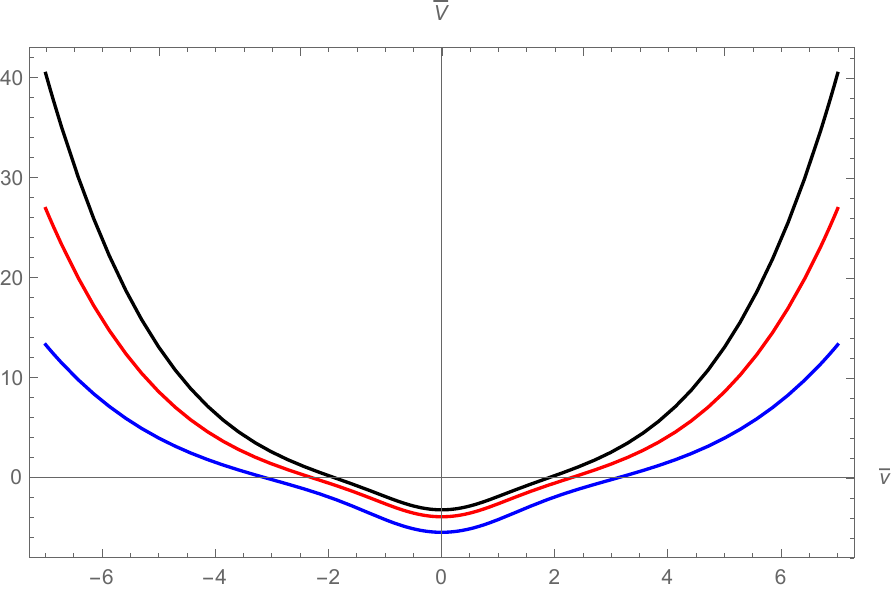}
  \includegraphics[width=7.8cm]{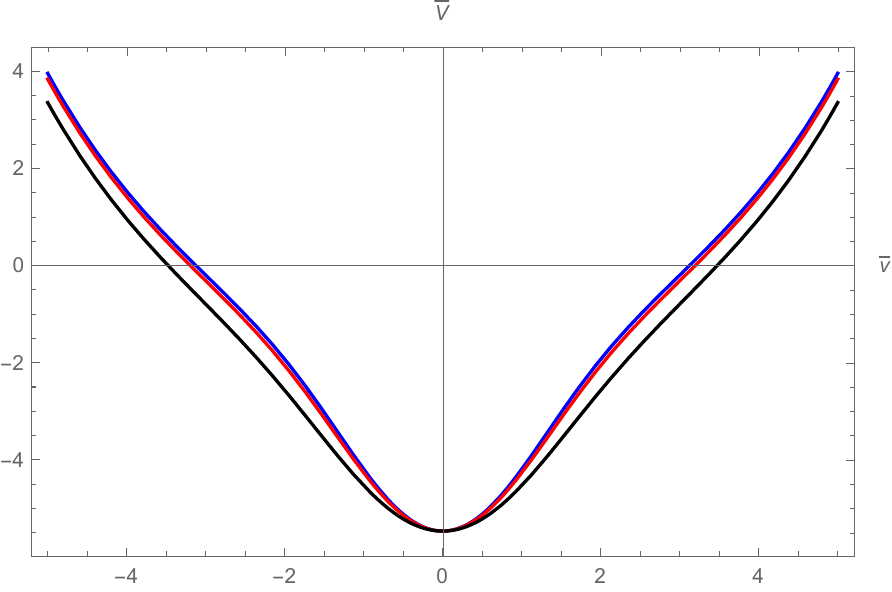}
 \caption{ \it \small Variation of the effective potential \ref{y73}, with respect to the background field $\Bar{v}$, for $\lambda\ensuremath{>}16 g^2$, \ref{y75}. Both left and right set of curves, correspond to massless cases.  For the left figure, the blue, red and black curves respectively correspond to $\lambda$ values $0.1$, $0.2$ and $0.3$, with  $g=0.01$. For the right figure, the blue, red and black curves respectively correspond to $g$ values $0.01$, $0.03$ and $0.06$, with $\lambda=0.1$. }
  \label{Vplot144}
\end{center}
\end{figure}

\begin{figure}[h!]
\begin{center}
 \includegraphics[width=7.8cm]{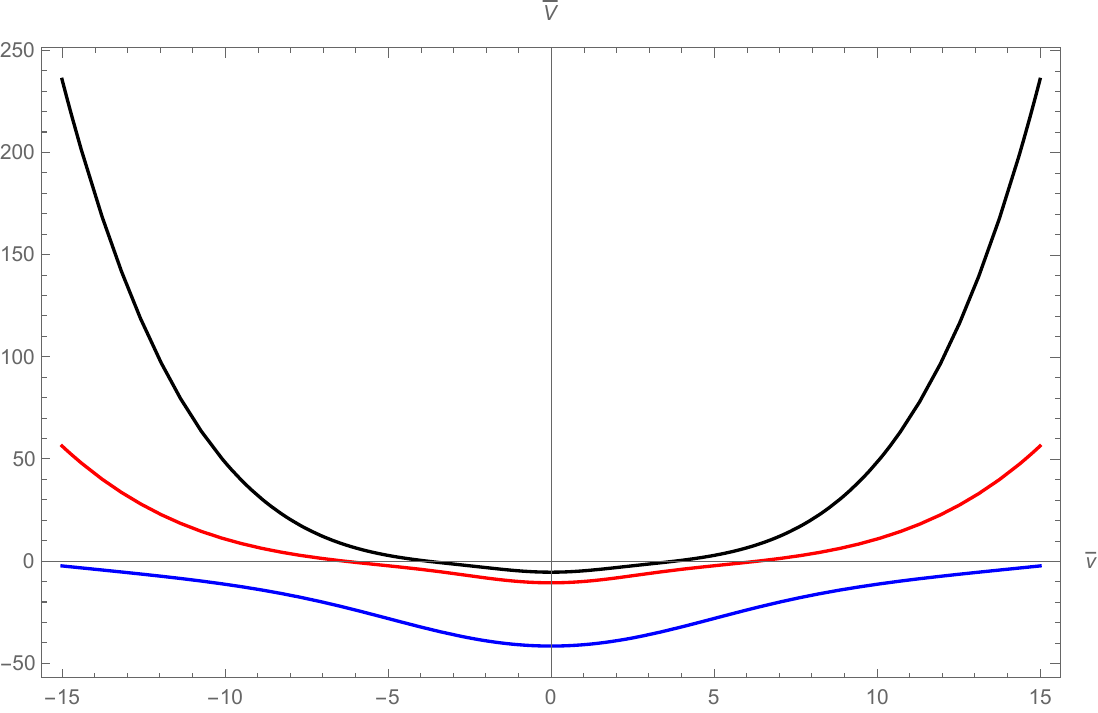}
\caption{ \it \small Variation of the effective potential \ref{y73}, with respect to the background field $\Bar{v}$, for $\lambda=16 g^2$, \ref{y75}. The blue, red and black curves respectively correspond to $g$ values $0.01$, $0.04$ and $0.08$. Note the qualitative difference with that of the $\lambda=16 g^2$ limit of the flat spacetime limit of the effective potential, \ref{y74}.}
 \label{Vplot145}
\end{center}
\end{figure}

\noindent
We have depicted the behaviour of the non-perturbative effective potential for massless fields, \ref{y73}, in \ref{Vplot144}, with $\lambda \ensuremath{>}16g^2$, \ref{y75}. Let us compare them with the pure scalar field theory results,  \ref{Vplot2}, \ref{Vplot1}. For a scalar with zero or positive rest mass squared, \ref{Vplot1} are rather similar to the tree level potential. \ref{Vplot144} on the other hand, shows behaviour qualitatively similar to that of \ref{Vplot2}, which corresponds to negative rest mass square for the scalar field in a pure $\phi^4$-theory. \ref{Vplot145} shows 
the                                                                                                                                                                                                                                                                                                                                                                                                                                                                                                                                                                                                                                                                     feature of the non-perturbative effective potential with $\lambda =16 g^2$. Note that for this equality, the $H \to 0$ limit of the effective potential is just the tree level quartic potential, which is very different from that of  \ref{Vplot145}. 
The above discussion shows that the non-perturbative quantum effects can bring in novel features in spacetimes like the de Sitter. We also recall that for very high energy phenomenon occurring at very small length scales one usually ignores gravity and spacetime curvature. Then the  comparison of our non-perturbative results with that of the $H \to 0$ limit also gives us one example of how quantum effects can be very different at short and large scales. Note also that  the spontaneous symmetry breaking feature of the flat spacetime (or short scale, sub-Hubble or UV) limit of the effective potential might lead to the formation of non-trivial vacuum structures like the topological defects. For \ref{Vplot133}, since we essentially have the breaking of the $v \to - v$ reflection symmetry, the corresponding defect will be the domain walls~\cite{Vilenkin:2000jqa}.  Also, as we have mentioned above, increasing the Yukawa coupling may make the potential unbounded from below for $\lambda \ensuremath{<} 16 g^2$. Hence until this limit is reached, increasing the $g$ value with a fixed $\lambda$ would deepen the minima of  \ref{Vplot133}. Thus we may expect that such defect formation would be aided by the Yukawa interaction within this limit. However we also note that,   since  the full large scale potential washes away the symmetry breaking pattern, it also  rules out any such defect formation at that scale. The bound $\lambda \gtrsim 16 g^2$ found above thus  ensures the stability of the vacuum at large scale. This once again strongly emphasises the energy scale dependence of the physical phenomenon we are considering. In other words, formation of topological defects for \ref{y73} is  possible only if the field is extremely high energetic, and hence confined to a radius of size much small compared to the Hubble horizon, but not otherwise.
\begin{figure}[h!]
\begin{center}
 \includegraphics[width=7.0cm]{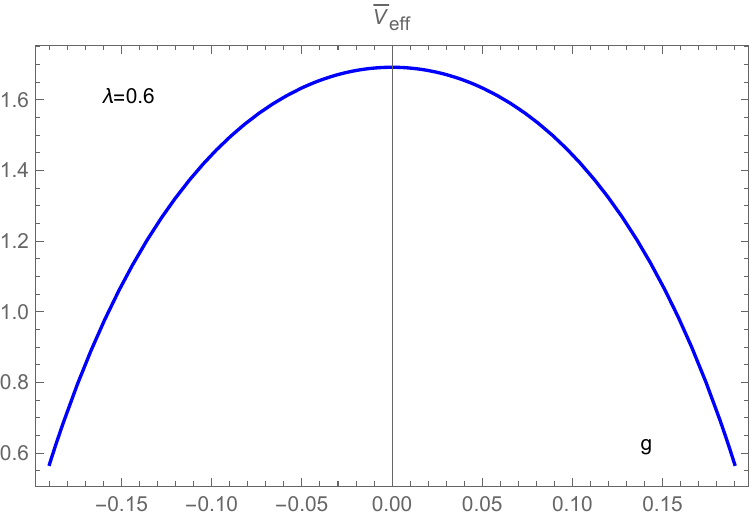}
 \includegraphics[width=7.0cm]{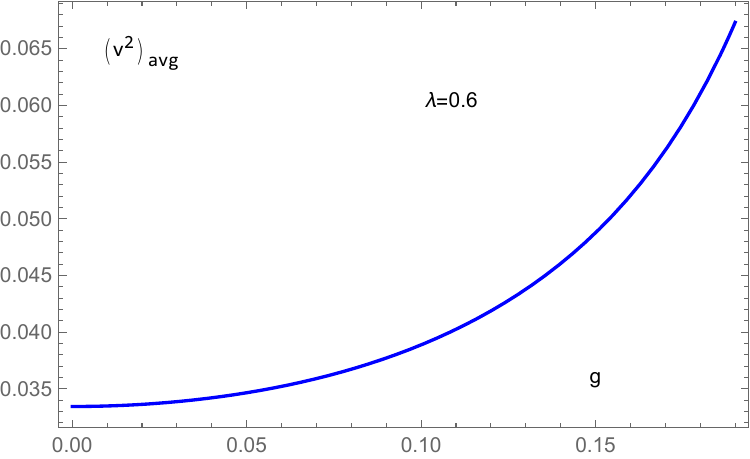}
\caption{ \it \small  (Left) Variation of the effective potential \ref{y73}, with respect to the Yukawa coupling $g$ for $v/H=2.0$. (Right)  $\langle v^2\rangle_{\rm stoch.}$ vs. $g$ plot (in the unit of $H^2$), assuming the background scalar becomes stochastic at late times.  $\lambda$ is taken to be $0.6$ in both plots.  Since the fermion mass is $M=gv$, the above plots effectively capture the variation with respect to $M$ as well. See main text  for discussion.}
 \label{Vplot146}
\end{center}
\end{figure}

 Let us also emphasise here the effect of the fermion mass ($M_{\rm dyn. eff}=M_0+gv$) in this context. Since we have explicitly focussed on the case $M_0=0$ here,  the fermion mass contribution comes as $g^2 v^2$ in \ref{y73}. This makes the effective potential and the action ${\cal Z}_2$ symmetric under the reflection $v\to -v$, as has been depicted in the figures. Retaining fermion rest mass however, would break this  reflection symmetry, \ref{y72}. This way the fermion mass can have qualitatively new effect on the correlation functions. We further note that the Yukawa term, including the fermion mass contributes to the effective dynamically generated  mass of the scalar field, \ref{y57a4}. This {\rm effectively} leads to the  negative rest mass squared like behaviour of \ref{Vplot2}. Note that for the rest mass squared positive, the behaviour of the effective potential is rather smooth,~\ref{Vplot1}. We have depicted the variation of the effective potential with respect to the Yukawa coupling in the first of~\ref{Vplot146}. Decrease of $V_{\rm eff. loc}$ with increasing $g$-value is consistent with the bound \ref{y75}. We will further come to this issue in the following Section.

 Finally, a possible caveat to our above analysis, as we have mentioned it earlier as well, is the estimation of the value of the secular logarithm numerically, rather than employing any formal resummation procedure (cf., discussion below \ref{y52a}). We wish to attempt now  assigning a formal, non-perturbative  value to this $\ln a$, should we {\it agree} to consider an independent equation for resummation, other than the ones that naturally arise in the 2PI formalism, and thereby simultaneously using two different methods. We adopt the renormalisation group inspired autonomous equation for resummation of de Sitter secular logarithms proposed a couple of years ago in~\cite{Kamenshchik:2020yyn}. This method was applied recently in~\cite{Bhattacharya:2023yhx}, in the context of the $\phi^4$-Yukawa theory as follows. At late times, and under the slow roll condition, the vacuum expectation value of the coincident two point scalar propagator satisfies at the leading order,
 \be
 \frac{d\langle \phi^2 \rangle }{d {\cal N}}= \frac{H^2}{4\pi^2} - \frac{2}{3H^2} \left( \frac{\lambda}{6}\langle \phi^4\rangle + g \langle \bar{\psi}\psi \phi \rangle\right),
 \label{add1}
 \ee
 where ${\cal N}= \ln a=Ht$ is the number of de Sitter e-foldings. We will take $\langle \phi^4\rangle= 3\langle \phi^2\rangle^2$ at the leading order in the Hartree approximation. Note that there is no explicit appearance of dynamical mass in \ref{add1}. This is because in this method the same is generated via $\langle \phi^2 \rangle$.  Note next that the secular logarithm term in \ref{y71} arises from the vacuum expectation value of the {\it local } part of the Yukawa term, $\bar{\psi}\psi \phi $, \ref{y45}.  \ref{add1} thus suggests that this secular logarithm  can be related to $\langle \phi^2 \rangle$. It also suggests that for an equilibrium state to exist ($d\langle \phi^2 \rangle/d {\cal N}=0$), the $\ln a$ term cannot be arbitrarily large.  Following~\cite{Kamenshchik:2020yyn}, we next construct the autonomous and non-perturbative equation for $\langle \phi^2 \rangle$, by eliminating the $\ln a$ terms. It turns out that in the equilibrium, the consistent, non-perturbative value associated with the secular logarithm in question  is given by~\cite{Bhattacharya:2023yhx},
 \be
\ln a \to  \frac {2\sqrt{3} \pi}{\sqrt{\lambda + 6 g^2}},
 \label{add2}
 \ee
 which we may substitute in \ref{y71}. In other words, we are basically replacing the $\ln a$ term by the non-perturbative value of $\langle \phi^2\rangle$, i.e. the scalar vacuum bubble. It is clear that this will bring in no qualitative difference to the result we found earlier by the crude estimation of  $\ln a$. However note that in this analysis, we have ignored the background scalar field $v$, and hence the conclusion holds good as long as $\bar{v}=v/H \ll 1$. We may next include $v$  as a rest mass term, $m_0^2 \langle \phi^2 \rangle $ ($m_0^2=\lambda v^2/2$), on the right hand side of \ref{add1}. Constructing the autonomous equation as earlier, it is easy to see that  we have an additional  $\lambda \bar{v}^2/2$ term added with $\lambda + 6 g^2$ in the denominator of  \ref{add2}. Thus for $\bar{v} \gg 1$, the $\ln a $ term in \ref{y71} will be suppressed compared to the $f_{\rm fin}$ term. Once again this is qualitatively similar to our earlier result, as  in the curly bracket of  \ref{y72}, the term multiplied with $0.7$ (originating from $\ln a$ following our numerical estimation) is subleading compared to the $f_{\rm fin}$ term.

Another rigorous approach to resum $\langle \bar{\psi}\psi \phi \rangle$ would be  trying to construct the 3PI effective action generating the three point and three particle irreducible effective action, generating a Schwinger-Dyson-like equation for $\langle \bar{\psi}\psi \phi \rangle$. Even though seemingly an extremely formidable task in de Sitter, it seems to be worth attempting.

\section{Conclusions}\label{S4}
In this paper we have obtained the local part of the two loop effective action for the $\phi^4$-Yukawa theory via non-perturbative 2PI effective action formalism. In \ref{S31}, we briefly sketch the derivation for the pure $\phi^4$ theory. In \ref{S32}, we discuss the inclusion of the Yukawa interaction. This paper is an extension of the previous works on the scalar field theory in the two loop Hartree approximation~\cite{Arai:2012sh, Arai:2013jna, LopezNacir:2013alw}. The Schwinger-Dyson equations satisfied by the scalar and fermion propagators have been constructed and various non-trivial non-perturbative counterterms has been found out explicitly, e.g. \ref{y50}, \ref{y57a1}. The dynamically generated masses of the scalar and the fermions have also been computed. We have pointed out non-trivial feature associated with cancellation of an overlapping divergence  in the effective action  (cf., discussion below \ref{y57}, \ref{y65}), absent in the standard 1PI perturbative formalism. \ref{y71}  is the renormalised result thus found. This contains a secular logarithm, $\ln a$, which is  absent in the scalar sector.  We  also have  shown briefly that in the perturbative technique,   the finite part actually contains a quadratic power of $\ln a$, \ref{E}, owing to the different renormalisation procedure.
We next explicitly investigate the behaviour of the effective potential for  zero rest mass cases. For the effective potential to be bounded from below, we find the constraint, $\lambda \gtrsim 16 g^2$, \ref{y75}. \ref{Vplot144} and \ref{Vplot145} shows the variation of the effective potential with respect to the background field. We note in particular the qualitative similarity of these plots with that of \ref{Vplot2}, which corresponds to {\it negative }rest mass squared of the scalar field. Putting things together, the present work perhaps shows the non-triviality of the non-perturbative effects in a spacetime like de Sitter. The discussion towards the end of the preceding section also shows how the results of the curved and flat spacetimes can be qualitatively very different.

We have also computed the 2PI two and three loop vacuum graphs for $\phi^4$ theory towards the end of \ref{S31} and \ref{A}, for  a massless and minimally coupled quantum scalar with {\it vanishing} background field. The background field term ($\lambda v^2/2$) acts like an effective mass term in the propagator. In the absence of it the `purely massless' scalar's propagator behaves in a qualitatively different manner  in de Sitter from that of a massive scalar, no matter how tiny its mass is~\cite{Brunier:2004sb}. We have computed in this particular case  both the local and leading non-local parts of the vacuum loops using the Schwinger-Keldysh formalism. To the best of our knowledge, diagram topology like \ref{fig3} where product of four propagators are present,  has not been attempted before in de Sitter. We believe the above results are interesting in their own right. \\

Let us now come to the issue of possible phenomenological implications of our effective potential, \ref{y73}. Following~\cite{Miao:2006pn},  we may treat the background field $v$ to be stochastic at late times, and compute the inflationary correlation functions. By computing the two or higher point correlators in the spatial momentum space and the associated power spectrum and non-Gaussianity, we may reasonably hope to put some bound on the Yukawa coupling, for example,  via the observed constraint on the spectral index, $1-n_s \lesssim 0.04$. As a warm up, let us consider here computing the {\it stochastic} average $\langle v^2(x)\rangle$~\cite{Starobinsky:1994bd},
$$\langle v^2(x)\rangle_{\rm stoch.} = N^{-1} \int_{-\infty}^{\infty} dv \ v^2  e^{-8\pi V_{\rm eff}(v)/3}$$
where $N= \int_{-\infty}^{\infty} dv \  e^{-8\pi V_{\rm eff}(v)/3}$ is the normalisation. The above has been plotted in the second of \ref{Vplot146}. Since the fermion mass is proportional to the Yukawa coupling,  $M_{\rm dyn. eff}=gv$, the plot qualitatively shows the possible effect of the fermion effective mass onto the two point correlation function.
Perhaps this might be an interesting result indicating that the Yukawa coupling makes the corresponding power spectrum blue tilted. However, a more realistic such computation should involve a quasi-de Sitter metric and associated slow roll parameters. In this case, we must treat the scalar field as the inflaton.

 Another interesting thing to look into would be the post inflationary reheating phenomenon with this effective potential. The dynamical mass residing in the effective potential contains the inflationary Hubble rate. At the end of the inflation, this Hubble rate would decay reducing the value of the dynamical mass and thereby releasing a huge amount of rest mass energy. Could this energy  reheat (or at least contribute to some extent in reheating) the universe? Certainly, this would require a coupling between the scalar and the photons. One can also look into the $O(N)$ symmetric scalar field theory without the Yukawa interaction for this purpose.   Finally we note that, we have focussed only on the local part of the effective potential. How do we compute the non-local parts? This involves solving the Schwinger-Dyson equations with non-local self energy contributions, and we need to use the Schwinger-Keldysh formalism. For scalar field theory some asymptotic analysis can be seen in~\cite{Youssef:2013by, Bhattacharya:2023xvd} with the two loop non-local self energy. Most importantly, how does this non-local part affect the correlation functions and the phenomenology of the spectral index? Can this part affect the small scale scenario like the Jean's instability during a gravitational collapse? All these seem to be  challenging as well as important tasks, and we hope to come back to at least some of them in the near future.

\bigskip

\section*{Acknowledgements} KR's research was partly supported by the fellowship from Council of Scientific and Industrial Research, Government of India (File No. 09/096(0987)/2019-EMR-I). The authors would like to sincerely acknowledge anonymous referee for many useful comments and suggestions. 

\bigskip
\appendix
\labelformat{section}{Appendix #1} 
\section{Three loop 2PI vacuum diagram at ${\cal O}(\lambda^2)$}\label{A}

In this appendix we wish to compute the three loop 2PI vacuum diagram (\ref{fig3}). We shall ignore any rest mass or background field term, and hence treat the scalar to be {\it purely} massless and minimally coupled.  We wish to compute both local and non-local contributions from \ref{fig3}, and hence we need to use the Schwinger-Keldysh, or in-in formalism. Let us first very briefly review this formalism, refering our reader to~\cite{Onemli:2002hr, Hu,  Adshead} for detail.

We recall that the standard time ordered functional integral representation of the standard in-out matrix elements for an observable of the field $A[\phi]$ read
\begin{eqnarray}
\langle \phi | T A[\phi] | \psi \rangle = \int {\cal D}\phi \ e^{i \int_{t_i}^{t_f} \sqrt{-g}d^d x {\cal L}[\phi]} \Phi^{\star}[\phi(t_f)] A[\phi] \Psi[\phi(t_i)]
\label{A1a}
\end{eqnarray}
where $|\phi\rangle $, $|\psi\rangle$ are field base-kets,  $\Phi[\phi]$, $\Psi[\phi]$ are wave functionals. The above matrix elements are well defined only if the asymptotic states are stable. However in a dynamical background such as the de Sitter, the initial vacuum state is not stable owing to the particle pair creation issues. Also, in such backgrounds, the interaction cannot be turned on and off and it should be omnipresent. In such non-equilibrium or dynamical scenario, one instead needs to resort  to the Schwinger-Keldysh formalism in order to compute any expectation value meaningfully.  

In order to introduce the in-in formalism, we write from \ref{A1a} for the anti-time ordering
\begin{eqnarray}
\langle \psi | \bar{T} A[\phi] | \phi \rangle = \left(\langle \phi | \bar{T} A[\phi] | \psi \rangle\right)^{\star} =\int {\cal D}\phi \ e^{-i \int_{t_i}^{t_f} \sqrt{-g}d^d x {\cal L}[\phi]} \Phi[\phi(t_i)] A[\phi] \Psi^{\star}[\phi(t_f)]
\label{A2a}
\end{eqnarray}
We have the completeness relationship on the final hypersurface at $t=t_f$, 
\begin{eqnarray}
\int {\cal D}\phi\ \Phi[\phi_-(t_f)]\Phi[\phi_+(t_f)]= \delta (\phi_+(t_f)-\phi_-(t_f))
\label{A3a}
\end{eqnarray}
From \ref{A1a}, \ref{A2a} and using \ref{A3a}, we have the in-in matrix elements
\begin{eqnarray}
\langle \psi | \bar{T} (A) T (A) | \psi \rangle = \int {\cal D}\phi_+ {\cal D}\phi_- \delta(\phi_+(t_f)-\phi_-(t_f))  e^{i \int_{t_i}^{t_f} \sqrt{-g}d^d x ({\cal L}[\phi_+] -{\cal L}[\phi_-])} \Psi^{\star}[\phi_-(t_i)]A[\phi_-] A[\phi_+] \Psi[\phi_+(t_i)]
\label{A4a}
\end{eqnarray}
Note that the field $\phi_+$ makes the forward time evolution, whereas $\phi_-$ makes backward time evolution.  We will take $|\psi\rangle$ to be the initial Bunch-Davies vacuum for a massless and minimal scalar field in de Sitter, $\nabla^2 \phi=0$, corresponding to the mode function 
\begin{eqnarray}
u_{k}(\eta)= \frac{H\eta}{\sqrt{2k}}\left(1-\frac{i}{k\eta} \right) e^{-ik\eta}
\label{A5a}
\end{eqnarray}
where $k$ is the modulus of the spatial momentum, $\vec{k}$.\\

\noindent
The three loop contribution to the 2PI effective action then reads (the other three loop contribution is a connected three bubble, which is not 2PI),
\be
i\Gamma_2^{3-{\rm loop}}(\lambda^2) = \frac{i \lambda^2}{48} \int (aa')^d d^d x d^d x' \left[i \Delta_{++}^4(x,x')- i\Delta_{+-}^4(x,x')\right]
\label{A5}
\ee
where $i \Delta_{++}(x,x')$ and $i\Delta_{+-}(x,x')= \langle \phi_+(x) \phi_-(x')\rangle$ are respectively the free Feynman propagator and the negative frequency Wightman function, \ref{S2}. Recall also that in our notation, $i\Delta^n \equiv (i\Delta)^n$.

Generically, from  \ref{props1}, \ref{props2}, we have the different segments for the fourth power of the propagators,
\begin{eqnarray}
i\Delta^4(x,x')= A^4+4A^3 B + 6A^2B^2 + 4 A B^3 + B^4 + 4A^3 C,
\label{A1}
\end{eqnarray}
 Note that we have omitted some terms containing $C(x,x')$. This is because the same is vanishing for either  $\e \to 0$ or  $\Delta x^2\to 0$, or both, \ref{props2}. Also, the term $B(x,x')$ is finite as $\e\to 0$. Thus first of all, all terms containing only $B$ and $C$ drop out.  Let us now consider the term $A^2 B C$. We can see from \ref{A3}, \ref{A4} below that this term behaves as $\sim \delta^d(x-x') B(x,x') C(x,x')/\e$, which vanishes in the coincidence limit due to the $\delta$-function, while integrating in \ref{A5}. Likewise all the other terms containing $A,\ B$ and $C$ vanish. It was pointed out earlier in~\cite{Brunier:2004sb} that in order to have any non-vanishing contribution associated with $C(x,x')$, it must be multiplied with cubic or higher power of $A$. We will show below that the $A^3 C$ term  appearing in \ref{A1} contains no UV divergence.  

 We next compute
\begin{eqnarray}
&& A^4 = \frac{\Gamma^4(1-\e/2)}{2^8 \pi^{8-2\e}}\frac{(aa')^{-4+2\e}}{\Delta x^{8-4\e}} \nonumber\\
&& A^3 B = \frac{H^{2-\e}\Gamma^2(1-\e/2)\Gamma(2-\e)(aa')^{-3+3\e/2}}{2^{9-\e}\pi^{8-2\e}}\left[- \frac{2\Gamma(3-\e/2)\Gamma(2-\e/2)(aa'H^2/4)^{\e/2}}{\e\Gamma(3-\e)} \frac{1}{\Delta x^{6-4\e}} +\left(\frac{2}{\e}+\ln (aa') \right) \frac{1}{\Delta x^{6-3\e}} \right] \nonumber\\
&&A^2B^2 = \frac{H^{4-2\e}\Gamma^2(3-\e)(aa')^{-2+\e}}{2^{12-2\e}\pi^{8-2\e}(1-\e)} \left[\frac{2^{2-2\e} H^{2\e}\Gamma^2(3-\e/2)\Gamma^2(2-\e/2)}{\e^2 \Gamma^2(3-\e)}\frac{(aa')^{\e}}{\Delta x^{4-4\e}} +\frac{(\ln(aa')+2/\e)^2}{\Delta x^{4-2\e}} \right. \nonumber\\&&\left.
-\frac{2^{2-\e}H^{\e} \Gamma(3-\e/2)\Gamma(2-\e/2)}{\e \Gamma(3-\e)} \frac{(aa')^{\e/2}(\ln(aa')+2/\e)}{\Delta x^{4-3\e}}\right] \nonumber\\&& AB^3= -\frac{H^6}{2^{11}\pi^8 aa'} \frac{\ln^3 \sqrt{e}H^2 \Delta x^2/4}{\Delta x^2}\nonumber\\&&
B^4 = \frac{H^8}{2^{12}\pi^8} \ln^4 \frac{\sqrt{e} H^2 \Delta x^2}{4}
\nonumber\\&& A^3 C= \frac{\Gamma^3(1-\e/2)}{2^6 \pi^{6-3\e/2}} \frac{(aa')^{-3+3\e/2} C(x,x')}{\Delta x^{6-3\e}}
\label{A2}
\end{eqnarray}
Following~\cite{Brunier:2004sb}, we next note for $d=4-\e$,
\begin{eqnarray}
&& \frac{1}{\Delta x^{8-4\e}}= \frac{1}{2(3-2\e)(4-3\e)} \p^2 \frac{1}{\Delta x^{6-4\e}} = \frac{1}{2^2(3-2\e)(4-3\e)(2-2\e)(2-3\e)} \p^4 \frac{1}{\Delta x^{4-4\e}}\nonumber\\&&
=- \frac{1}{2^3 3  (3-2\e) (4-3\e)(2-2\e)(2-3\e)(1-2\e)\e} \p^6 \frac{1}{\Delta x^{2-4\e}} \nonumber\\&&
\frac{1}{\Delta x^{6-4\e}}= - \frac{1}{2^2 \cdot 3  (2-2\e)(2-3\e)(1-2\e)\e} \p^4 \frac{1}{\Delta x^{2-4\e}}\nonumber\\&&
\frac{1}{\Delta x^{6-3\e}}= - \frac{1}{2^3   (2-2\e)(2-3\e/2)(1-3\e/2)\e} \p^4 \frac{1}{\Delta x^{2-3\e}}\nonumber\\&&
\frac{1}{\Delta x^{4-4\e}}= - \frac{1}{2\cdot 3 (1-2\e)\e} \p^2 \frac{1}{\Delta x^{2-4\e}}\nonumber\\&&
\frac{1}{\Delta x^{4-3\e}}= - \frac{1}{2^2 (1-3\e/2)\e} \p^2 \frac{1}{\Delta x^{2-3\e}} \nonumber\\&&
\frac{1}{\Delta x^{4-2\e}}= - \frac{1}{2 (1-\e)\e} \p^2 \frac{1}{\Delta x^{2-2\e}}
\label{A3}
\end{eqnarray}
We also note from \ref{props3'} that 
$$\p^2 \frac{1}{\Delta x_{++}^{2-\e}} =  \frac{4i \pi^{2-\e/2}}{\Gamma(1-\e/2)} \delta^d (x-x')$$
Thus from \ref{A3}, we have 
\begin{eqnarray}
&& \frac{1}{\Delta x^{8-4\e}_{++}}
=- \frac{i \mu^{-3\e} \pi^{2-\e/2}}{6  (3-2\e) (4-3\e)(2-2\e)(2-3\e)(1-2\e)\e \Gamma(1-\e/2)}   \p^4 \delta^d (x-x') -\frac{1}{3\times 2^8}\p^6 \frac{\ln \mu^2 \Delta x_{++}^2}{\Delta x_{++}^2} \nonumber\\&&
\frac{1}{\Delta x^{6-4\e}_{++}}= -\frac{i \mu^{-3\e} \pi^{2-\e/2}}{3(2-2\e)(2-3\e)(1-2\e)\e \Gamma(1-\e/2)}\p^2 \delta^d (x-x') -\frac{1}{ 2^5} \p^4 \frac{\ln \mu^2 \Delta x_{++}^2}{\Delta x_{++}^2}\nonumber\\&&
\frac{1}{\Delta x^{6-3\e}_{++}} =-\frac{i \mu^{-2\e} \pi^{2-\e/2}}{2(2-2\e)(2-3\e/2)(1-3\e/2)\e \Gamma(1-\e/2)}\p^2 \delta^d (x-x') -\frac{1}{ 2^5} \p^4 \frac{\ln \mu^2 \Delta x_{++}^2}{\Delta x_{++}^2}\nonumber\\&&
\frac{1}{\Delta x^{4-4\e}_{++}}= -\frac{2i \mu^{-3\e} \pi^{2-\e/2}}{3(1-2\e)\e \Gamma(1-\e/2)} \delta^d (x-x') -\frac{1}{2^2} \p^2 \frac{\ln \mu^2 \Delta x_{++}^2}{\Delta x_{++}^2}\nonumber\\&&
\frac{1}{\Delta x^{4-3\e}_{++}}= -\frac{i \mu^{-2\e} \pi^{2-\e/2}}{(1-3\e/2)\e \Gamma(1-\e/2)} \delta^d (x-x') -\frac{1}{2^2} \p^2 \frac{\ln \mu^2 \Delta x_{++}^2}{\Delta x_{++}^2}\nonumber\\&&
\frac{1}{\Delta x^{4-2\e}_{++}}= -\frac{2i \mu^{-\e} \pi^{2-\e/2}}{(1-\e)\e \Gamma(1-\e/2)} \delta^d (x-x') -\frac{1}{2^2} \p^2 \frac{\ln \mu^2 \Delta x_{++}^2}{\Delta x_{++}^2} \nonumber\\&&
\frac{1}{\Delta x^{6-2\e}_{++}}= - \frac{i \mu^{-\e}\pi^{2-\e/2}}{(1-\e)(2-\e)^2\Gamma(1-\e/2)\e} \p^2 \delta^d(x-x')-\frac{1}{2^5} \p^4 \frac{\ln \mu^2 \Delta x_{++}^2}{\Delta x_{++}^2}
\label{A4}
\end{eqnarray}
where $\mu$ is an arbitrary  scale having the dimension of mass. Note in the above expressions, that the terms associated with the $\delta$-functions are local and carry divergences, whereas the logarithms are non-local. The $\delta$-functions arise from the $(|\eta-\eta'|\mp i\e)$ terms appearing in $\Delta^2_{++}(x,x')$ or $\Delta^2_{--}(x,x')$, \ref{props3'}. Thus   terms containing $\Delta x^2_{\pm \mp}$ cannot yield any local contribution or divergences, but they contribute to the deep infrared.\\

\noindent
We wish to first compute below the local contribution in \ref{A5}, arising from the Feynman propagator ($++$). This will require renormalisation, following which we will obtain finite local contribution. After this, we shall compute the aforementioned  deep infrared, non-local contribution separately. 
Thus for  $i\Delta^4_{++}(x,x')$  in \ref{A5}, the contribution from the $A_{++}^4$ term, \ref{A2}, reads after using \ref{A4},
\begin{eqnarray}
&& \frac{i\lambda^2 \Gamma^4(1-\e/2)}{2^{12}\times 3 \pi^{8-2\e}} \int d^d x d^d x' (aa')^d  \frac{(aa')^{-4+2\e}}{\Delta x^{8-4\e}_{++}} \nonumber\\&&=\frac{\mu^{-3\e}\lambda^2 \Gamma^3(1-\e/2)}{2^{13}\times 9 \pi^{6-3\e/2} (3-2\e)(4-3\e)(2-2\e)(2-3\e)(1-2\e)\e} \int d^d x d^d x' (aa')^{\e} \p^4 \delta^d (x-x') 
=\frac{\lambda^2 H^4}{2^{15}\times 9\pi^6} \int d^4x a^4 \nonumber\\
\label{A6}
\end{eqnarray}
Note that the above term is free of any ultraviolet divergence.\\

\noindent
We next consider the $4A_{++}^3B_{++}$ term in \ref{A5}, \ref{A2}. Using the second and the third of \ref{A4}, we have for the local contribution
\begin{eqnarray}
&& \frac{i \lambda^2 H^{2-\e} \Gamma^2(1-\e/2) \Gamma(2-\e)}{2^{11-\e} \cdot 3 \pi^{8-2\e}} \int d^d x d^d x'(aa')^{d-3+3\e/2}\left[- \frac{2\Gamma(3-\e/2)\Gamma(2-\e/2)(aa'H^2/4)^{\e/2}}{\e\Gamma(3-\e)} \frac{1}{\Delta x^{6-4\e}_{++}} \right. \nonumber\\&&\left.+\left(\frac{2}{\e}+\ln (aa') \right) \frac{1}{\Delta x^{6-3\e}_{++}} \right]\nonumber\\&&
= \frac{i \lambda^2 H^{2-\e} \Gamma^2(1-\e/2) \Gamma(2-\e)}{2^{11-\e} \cdot 3 \pi^{8-2\e}} \int d^d x d^d x'(aa')^{d-3+3\e/2} \left[ \frac{i\mu^{-3\e} H^{\e} 2^{-\e} \pi^{2-\e/2} \Gamma(3-\e/2)  (aa')^{\e/2}}{6 \Gamma(2-\e)(1-\e)(2-3\e)(1-2\e)\e^2} \p^2 \delta^d(x-x') \right. \nonumber\\&& \left. 
-\left(\frac{2}{\e}+\ln (aa') \right) \frac{i \mu^{-2\e} \pi^{2-\e/2}}{2(2-2\e)(2-3\e/2)(1-3\e/2)\e \Gamma(1-\e/2)}\p^2 \delta^d (x-x')
 \right] \nonumber\\&&
 =- \frac{\lambda^2 \mu^{-3\e} H^2 \Gamma^2(1-\e/2)\Gamma(3-\e/2)}{2^{12}\times 9\pi^{6-3\e/2} (1-\e)(2-3\e)(1-2\e)\e^2} \int d^d x d^d x'(aa')^{1+\e}\p^2 \delta^d (x-x')  \nonumber\\&&
 + \frac{\lambda^2 \mu^{-2\e} H^{2-\e} \Gamma(1-\e/2)\Gamma(1-\e)}{2^{13-\e} \cdot 3 \pi^{6-3\e/2}(2-3\e/2)(1-3\e/2)\e} \int d^d x d^d x'(aa')^{1+\e/2}\left(\frac{2}{\e}+\ln (aa') \right) \p^2 \delta^d (x-x')\nonumber\\&&
 = \frac{\lambda^2 \mu^{-2\e} H^{4-\e}(1+\e)(3+2\e) \Gamma^2(1-\e/2)\Gamma(3-\e/2)}{2^{11}\cdot 9\pi^{6-3\e/2} (1-\e)(2-3\e)(1-2\e)\e^2}\left(\frac{\mu}{H} \right)^{-\e} \int d^d x a^d \left(1+3\e \ln a +\frac{9\e^2}{2} \ln^2 a +{\cal O}(\e^3) \right)\nonumber\\&& -\frac{\lambda^2 \mu^{-2\e} H^{4-\e} (2+\e)(3+\e) \Gamma(1-\e/2)\Gamma(1-\e)}{2^{12-\e} \cdot 3 \pi^{6-3\e/2}(2-3\e/2)(1-3\e/2)\e^2} \int d^d x a^d \left(1+ 2\e \ln a + 2\e^2 \ln^2 a  +{\cal O}(\e^3) \right) \nonumber\\&&
 -\frac{\lambda^2 \mu^{-2\e} H^{4-\e} \Gamma(1-\e/2)\Gamma(1-\e)}{2^{12-\e} \cdot 3 \pi^{6-3\e/2}(2-3\e/2)(1-3\e/2)\e} \int d^d x a^d \left[ (2+\e)(3+\e)\ln a (1+ 2\e \ln a) + (5+ 2\e)(1+ 2\e \ln a ) +{\cal O}(\e^2) \right]\nonumber\\
\label{A7}
\end{eqnarray}

\noindent
Let us now consider the $6A^2B^2$ term in \ref{A2}. Using the last three equations of \ref{A4}, we have for the local contribution
\begin{eqnarray}
&& \frac{i \lambda^2 H^{4-2\e} \Gamma^2(3-\e) }{2^{15-2\e} \pi^{8-2\e}(1-\e)} \int d^d x d^d x'(aa')^{d-2+\e} \left[\frac{2^{2-2\e} H^{2\e}\Gamma^2(3-\e/2)\Gamma^2(2-\e/2)}{\e^2 \Gamma^2(3-\e)}\frac{(aa')^{\e}}{\Delta x^{4-4\e}_{++}} +\frac{(\ln(aa')+2/\e)^2}{\Delta x^{4-2\e}_{++}} \right. \nonumber\\&&\left.
-\frac{2^{2-\e}H^{\e} \Gamma(3-\e/2)\Gamma(2-\e/2)}{\e \Gamma(3-\e)} \frac{(aa')^{\e/2}(\ln(aa')+2/\e)}{\Delta x^{4-3\e}_{++}}\right]
\nonumber\\&&      =   \frac{\lambda^2 \mu^{-2\e} H^{4-\e} \Gamma^2(3-\e/2)\Gamma^2(2-\e/2)}{2^{12}\cdot 3 \pi^{6-3\e/2}  (1-\e)(1-2\e)  \Gamma(1-\e/2)\e^3} \int d^d x a^d \left(1+3\e \ln a + \frac{9\e^2}{2}\ln^2 a + \frac{9\e^3}{2}\ln^3 a +{\cal O}(\e^4) \right) \nonumber\\&&
+ \frac{\lambda^2 \mu^{-2\e}H^{4-2\e}\Gamma^2(3-\e)}{2^{12-2\e} \pi^{6-3\e/2}(1-\e)^2 \Gamma(1-\e/2)\e^3} \int d^d x a^d \left(1+ 3\e \ln a +\frac{7\e^2}{2} \ln^2 a + \frac{13 \e^3}{6} \ln^3 a +{\cal O}(\e^4) \right) \nonumber\\&&
- \frac{\lambda^2 \mu^{-2\e}H^{4-\e}\Gamma(3-\e) \Gamma(3-\e/2)\Gamma(2-\e/2)}{2^{12-\e} \pi^{6-3\e/2}(1-\e)(1-3\e/2) \Gamma(1-\e/2)\e^3} \int d^d x a^d \left(1+ 3\e \ln a +4\e^2 \ln^2 a + \frac{10 \e^3}{3} \ln^3 a +{\cal O}(\e^4) \right)
\label{A8}
\end{eqnarray}

\noindent
It can be shown   that the $AB^3$ and $B^4$ terms do not contribute to any local terms in the effective action. The only remaining term relevant for this purpose will be the $4A^3 C$ term in \ref{A2}, 
\begin{eqnarray}
&& \frac{i \lambda^2  \Gamma^3(1-\e/2) }{2^{8} \cdot 3 \pi^{6-3\e/2}} \int d^d x d^d x'(aa')^{d-3+3\e/2} \frac{C_{++}(x,x')}{\Delta x^{6-3\e}_{++}} \nonumber\\&&
=  \frac{\lambda^2 \mu^{-2\e} \Gamma^2(1-\e/2)}{2^{10} \cdot 3 \pi^{4-\e} (1-\e)(2-3\e/2)(1-3\e/2)\Gamma(3-\e/2)\e}  \int d^d x d^d x' (a a')^{d-3+3\e/2} C_{++}(x,x') \p^2 \delta^d (x-x')
\label{A9}
\end{eqnarray}
We integrate the above equation by parts. From \ref{props2} we  note that in $3-\e$ spatial dimensions,
\begin{eqnarray}
\p^2 \Delta x^2_{++} = \vec{\p}^2 |\vec{x} -\vec{x'}|^2 + \p_0^2 |\eta-\eta'|^2 = 2(3-\e) + 4 (\eta-\eta') \delta (\eta-\eta') +2 ({\rm sgn}(\eta-\eta'))^2
\label{A10}
\end{eqnarray}
where the sgn stands for the sign of $(\eta-\eta')$ and it is zero for $\eta=\eta'$. Putting things together now, \ref{A9} becomes 
\begin{eqnarray}
&& \frac{\lambda^2 \mu^{-2\e} H^{4-\e} \Gamma^2(1-\e/2)(1-\e/3)}{2^{15-\e}  \pi^{6-3\e/2} (1-\e)(2-3\e/2)(1-3\e/2)\Gamma(3-\e/2)\e} \left(\frac{\Gamma(4-\e)}{\Gamma(3-\e/2)} - \frac{\Gamma(4-\e/2)}{2^{\e}(1+\e/2)\Gamma(3)}\right) \int d^d x a^d a^{2\e}
\label{A11}
\end{eqnarray}
It is easy to check by expanding the Gamma functions within parenthesis that the above term is not divergent. Thus we may set $d=4$, $a^{2\e}=1$ above and hence the whole term can be absorbed in a finite cosmological constant counterterm, $\delta \Lambda/8\pi G$. Combining now \ref{A6}, \ref{A7}, \ref{A8} and \ref{A11}, we have the three loop, local contribution 
\begin{eqnarray}
&& \Gamma_{2,\ \rm loc}^{3-{\rm loop}}(\lambda^2) = \frac{\lambda^2 H^4}{2^{15}\pi^6}\left[\frac19 +\frac{1}{4\e}\left(\frac{\Gamma(4-\e)}{\Gamma(3-\e/2)} - \frac{\Gamma(4-\e/2)}{2^{\e}(1+\e/2)\Gamma(3)}\right)\right] \int d^4 x a^4 \nonumber\\&&
+\frac{\lambda^2 \mu^{-2\e} H^{4-\e}}{2^{11}\pi^{6-3\e/2} } \left[ \frac{(1+\e)(3+2\e) \Gamma^2(1-\e/2)\Gamma(3-\e/2)}{9 (1-\e)(2-3\e)(1-2\e)\e^2}\left(\frac{\mu}{H} \right)^{-\e} -\frac{ (2+\e)(3+\e) \Gamma(1-\e/2)\Gamma(1-\e)}{2^{1-\e} 3 (2-3\e/2)(1-3\e/2)\e^2}  \right. \nonumber\\&& \left. -\frac{ \Gamma(1-\e/2)\Gamma(1-\e)}{2^{1-\e} 3 (2-3\e/2)(1-3\e/2)\e} + \frac{ \Gamma^2(3-\e/2)\Gamma^2(2-\e/2)}{6  (1-\e)(1-2\e)  \Gamma(1-\e/2)\e^3}  + \frac{\Gamma^2(3-\e)}{2^{1-2\e}(1-\e)^2 \Gamma(1-\e/2)\e^3}\right. \nonumber\\&&\left. - \frac{\Gamma(3-\e) \Gamma(3-\e/2)\Gamma(2-\e/2)}{2^{1-\e} (1-\e)(1-3\e/2) \Gamma(1-\e/2)\e^3} \right] \int d^d x a^d \nonumber\\&&
+ \frac{\lambda^2 \mu^{-2\e} H^{4-\e}}{2^{11} \pi^{6-3\e/2}\e} \left[ \frac{(1+\e)(3+2\e)\Gamma^2(1-\e/2)\Gamma(3-\e/2)}{3(1-\e)(2-3\e)(1-2\e)} \left(\frac{\mu}{H} \right)^{-\e} - \frac{(2+\e)(3+\e)\Gamma(1-\e/2)\Gamma(1-\e)}{2^{-\e} 3(2-3\e/2)(1-3\e/2)}  \right. \nonumber\\&&\left. -  \frac{\Gamma(1-\e/2)\Gamma(1-\e)}{2^{-\e}(2-3\e/2)(1-3\e/2)} +  \frac{\Gamma^2(3-\e/2)\Gamma^2(2-\e/2)}{2(1-\e)(1-2\e)\Gamma(1-\e/2)\e} + \frac{3\Gamma^2(3-\e)}{2^{1-2\e}(1-\e)^2\Gamma(1-\e/2)\e} \right. \nonumber\\&& \left.   - \frac{3\Gamma(3-\e)\Gamma(3-\e/2)\Gamma(2-\e/2)}{2^{1-\e}(1-\e)(1-3\e/2) \Gamma(1-\e/2)\e} \right]  \int d^d x a^d \ln a  + \frac{\lambda^2 \mu^{-2\e} H^{4-\e}}{2^{12} \pi^{6-3\e/2}\e}\left[\frac{3\Gamma^2(3-\e/2)\Gamma^2(2-\e/2)}{2(1-\e)(1-2\e)\Gamma(1-\e/2)}  \right. \nonumber\\&&\left. +\frac{7\Gamma^2(3-\e)}{2^{1-2\e}(1-\e)^2\Gamma(1-\e/2)}-\frac{4\Gamma(3-\e)\Gamma(3-\e/2)\Gamma(2-\e/2)}{2^{-\e}(1-\e)(1-3\e/2)\Gamma(1-\e/2)} \right]\int d^d x a^d \ln^2 a + \frac{\lambda^2 H^4}{2^{10}\cdot 3\pi^6} \int d^4 x a^4 \left(\ln^3 a + {\cal O}(\ln^2 a) \right)
\nonumber\\
\label{A12}
\end{eqnarray}

The divergence associated with the $\ln^2 a$ term can be absorbed by adding with the above the contribution from a quartic vertex counterterm
$$-\frac{\delta \lambda}{4!} \int a^d d^d x\langle \phi^4 \rangle = - \frac{\delta \lambda }{2^3}\int a^d d^d x \ i\Delta^2(x,x) = -\frac{\delta \lambda H^{4-2\e}\Gamma^2(2-\e)}{2^{7-2\e}\pi^{4-\e}\Gamma^2(1-\e/2)}\int a^d d^d x \left(\frac{1}{\e^2}+\frac{2\ln a}{\e}+\ln^2 a \right) $$
where we have used \ref{y6}. This leads to the choice,
\begin{eqnarray}
&& \delta \lambda = \frac{\mu^{-2\e}\lambda^2H^{\e}\Gamma^2(1-\e/2)}{2^{5+2\e}\pi^{2-\e/2}\Gamma^2(2-\e) \e} \left[\frac{3\Gamma^2(3-\e/2)\Gamma^2(2-\e/2)}{2(1-\e)(1-2\e)\Gamma(1-\e/2)}   +\frac{7\Gamma^2(3-\e)}{2^{1-2\e}(1-\e)^2\Gamma(1-\e/2)}-\frac{4\Gamma(3-\e)\Gamma(3-\e/2)\Gamma(2-\e/2)}{2^{-\e}(1-\e)(1-3\e/2)\Gamma(1-\e/2)} \right] \nonumber\\
\label{A16}
\end{eqnarray}
The divergence with the $\ln a$ term can be cancelled via  a mass renormalisation counterterm,
\be
 - \frac{\delta m^2}{2}\int a^d d^d x \langle \phi^2 \rangle  =  - \frac{\delta m^2 H^{2-\e}\Gamma(2-\e)}{2^{3-\e}\pi^{2-\e/2}\Gamma(1-\e/2)}\int a^d d^d x \left(\ln a +\frac{1}{\e} \right)
 \ee
with the choice 
\begin{eqnarray}
&& \delta m^2  = \frac{\delta \lambda H^{2-\e}\Gamma(2-\e)}{2^{3-\e}\pi^{2-\e/2}\Gamma(1-\e/2)\e} - \frac{\mu^{-2\e}\lambda^2 H^2 \Gamma(1-\e/2)}{2^{8+\e}\pi^{4-\e}\Gamma(2-\e)\e}\left[ \frac{(1+\e)(3+2\e)\Gamma^2(1-\e/2)\Gamma(3-\e/2)}{3(1-\e)(2-3\e)(1-2\e)} \left(\frac{\mu}{H} \right)^{-\e} \right. \nonumber\\&&\left. - \frac{(2+\e)(3+\e)\Gamma(1-\e/2)\Gamma(1-\e)}{2^{-\e}\cdot 3(2-3\e/2)(1-3\e/2)}   -  \frac{\Gamma(1-\e/2)\Gamma(1-\e)}{2^{-\e}(2-3\e/2)(1-3\e/2)} +  \frac{\Gamma^2(3-\e/2)\Gamma^2(2-\e/2)}{2(1-\e)(1-2\e)\Gamma(1-\e/2)\e}\right. \nonumber\\&& \left.  + \frac{3\Gamma^2(3-\e)}{2^{1-2\e}(1-\e)^2\Gamma(1-\e/2)\e}    - \frac{3\Gamma(3-\e)\Gamma(3-\e/2)\Gamma(2-\e/2)}{2^{1-\e}(1-\e)(1-3\e/2) \Gamma(1-\e/2)\e} \right] 
\label{A16'}
\end{eqnarray}
 The remaining constant terms can be absorbed in the renormalisation of the cosmological constant.  Thus at three loop we have the renormalised local expression for the vacuum graph, \ref{fig3} 
\begin{eqnarray}
&& i\Gamma_2^{3-{\rm loop}}\vert_{\rm loc., ren.}= \frac{\lambda^2 H^4}{2^{10}\times 3\pi^6}  \int a^4 d^4 x \left[ \ln^3 a + {\cal O}(\ln^2 a) \right]
\label{A17}
\end{eqnarray}

\noindent
Let us now come to the non-local part of $\Gamma_2^{3-{\rm loop}}$. This will yield  the leading deep infrared contribution of ${\cal O}(\ln^4 a) $. We shall use the in-in formalism to do this.  Now, instead of using the exact expressions of \ref{A4}, we wish to use the IR effective formalism described in e.g.~\cite{Bhattacharya:2022wjl}. We have from \ref{A5} appropriate for the in-in and non-local contribution
\be
i\Gamma_2^{3-{\rm loop}}(\lambda^2)_{\rm non-loc} = \frac{i \lambda^2}{48} \int (aa')^d d^d x d^d x' \left[i \Delta_{++}^4(x,x')- i\Delta_{+-}^4(x,x')\right]= \frac{i \lambda^2}{48} \int (aa')^4 d^4 x d^4 x' \left[i \Delta_{-+}^4(x,x')- i\Delta_{+-}^4(x,x')\right]
\label{A5'}
\ee
where we have used for the Feynman propagator 
$$
i\Delta_{++}(x,x') = \theta(\eta-\eta')i\Delta_{-+}(x,x') +\theta(\eta'-\eta)i\Delta_{+-}(x,x')
$$
We now rewrite \ref{A5'} in the spatial momentum space. In the deep IR, super-Hubble limit, we have $H\lesssim k \lesssim Ha$, where $k= |\vec{k}|$. 
\begin{eqnarray}
&& i\Gamma_2^{3-{\rm loop}}\vert_{\rm leading, IR}= \frac{i\lambda^2}{48}  \int a^4 d^4 x \int a'^4 d \eta' \frac{d^3\vec{k_1} d^3\vec{k_2}d^3\vec{k_3}}{(2\pi)^9} \left(i\Delta_{-+}(k_1,\eta,\eta') -{\rm c.c.}\right) \left(i\Delta_{-+}(k_2,\eta,\eta') +{\rm c.c.}\right)\nonumber\\&& \times \left(i\Delta_{-+}(k_3,\eta,\eta')i\Delta_{-+}(|{\vec k_1}+{\vec k_2}+{\vec k_3}|,\eta,\eta') +{\rm c.c.}\right)\Theta(Ha'-k_1)\Theta(Ha'-k_2)\Theta(Ha'-k_3)\Theta(Ha'-|{\vec k_1}+{\vec k_2}+{\vec k_3}|)\nonumber\\&&
\simeq \frac{\lambda^2 H^4}{2^{11}\times 9\pi^6} \int a^4 d^4 x \ln^4 a
\label{A18}
\end{eqnarray}
where we have used $\eta \gtrsim \eta'$, or $\eta$ as the final time, and for the Wightman function,
\begin{eqnarray}
&& i\Delta_{+-}(k, \eta, \eta')= \frac{H^2 \Theta(Ha'-k)\Theta(Ha-k)}{2k^3}\left(1- \frac{ik^3}{3H^3 a'^3} \right) =  (i\Delta_{-+}(k, \eta, \eta'))^{\star} 
\label{A19}
\end{eqnarray}
\ref{A18}, combined with \ref{A17} gives the renormalised, leading late time expression for $i\Gamma_2^{3-{\rm loop}}$ at ${\cal O}(\lambda^2)$, yielding \ref{y43''}.

\section{Propagators and their coincidence limits}\label{D}
The Green function for a scalar field is given by~\cite{Bunch:1978yq}
\be
 iG(x,x')= \frac{H^{2-\e}}{2^{4-\e}\pi^{2-\e/2}} \frac{\Gamma\left(\frac32+\nu-\frac{\e}{2}\right)\Gamma\left(\frac32-\nu -\frac{\e}{2}\right)}{\Gamma\left(2-\frac{\e}{2}\right)} {_2F_1} \left(\frac32+\nu-\frac{\e}{2}, \frac32-\nu-\frac{\e}{2}, 2-\frac{\e}{2}, 1-\frac{y}{4}\right) 
\label{d1}
\ee
where $\nu = ((d-1)^2/4- m^2/H^2)^{1/2}$ and the parameter $\e=0^+$ has been kept for regularisation purpose. Using the transformation formula for the hypergeometric function~\cite{Grad}, 
\be
 _2F_1(\alpha, \beta, \gamma, z) = \frac{\Gamma(\gamma)\Gamma(\gamma-\alpha-\beta)}{\Gamma(\gamma-\alpha)\Gamma(\gamma-\beta)} {_2F_1}(\alpha,\beta, \alpha+\beta-\gamma+1, 1-z) + (1-z)^{\gamma-\alpha-\beta} \frac{\Gamma(\gamma)\Gamma(\alpha+\beta-\gamma)}{\Gamma(\alpha)\gamma(\beta)}{_2F_1}(\gamma-\alpha,\gamma-\beta,\gamma- \alpha-\beta+1, 1-z),
\label{d2}
\ee
we have
\begin{eqnarray}
 &&{_2F_1} \left(\frac32+\nu-\frac{\e}{2}, \frac32-\nu-\frac{\e}{2}, 2-\frac{\e}{2}, 1-\frac{y}{4}\right) = \frac{\Gamma(2-\e/2)\Gamma(-1+\e/2)}{\Gamma(1/2-\nu)\Gamma(1/2+\nu)}{_2F_1} \left(\frac{3}{2}+\nu-\frac{\e}{2}, \frac{3}{2}-\nu-\frac{\e}{2}, 2-\frac{\e}{2}, \frac{y}{4}\right) \nonumber\\&&
 + \left(\frac{y}{4} \right)^{-1+\e/2} \frac{\Gamma(1-\e/2)\Gamma(2-\e/2)}{\Gamma(3/2+\nu-\e/2)\Gamma(3/2-\nu-\e/2)}{_2F_1} \left(\frac12-\nu, \frac12+\nu, \frac{\e}{2}, \frac{y}{4}\right)
\label{d3}
\end{eqnarray}
For our purpose of computing local parts of the vacuum diagrams, we wish to make an expansion of the Green function for small separation, $y$. We have 
\begin{eqnarray}
 &&iG(y\to 0) = \frac{\Gamma(1-\e/2)}{2^2\pi^{2-\e/2}}\frac{(aa')^{-1+\e/2}}{\Delta x^{2-\e}} + \frac{H^{2-\e}}{2^{4-\e}\pi^{2-\e/2}} \frac{\Gamma(3/2+\nu-\e/2)\Gamma(3/2-\nu-\e/2)}{\Gamma(1/2-\nu)\Gamma(1/2+\nu)} \Gamma(-1+\e/2) +{\cal O}(y)\nonumber\\
 && = \frac{\Gamma(1-\e/2)}{2^2\pi^{2-\e/2}}\frac{(aa')^{-1+\e/2}}{\Delta x^{2-\e}} + \frac{H^{2-\e}}{2^{3-\e}\pi^{2-\e/2}}\left[\frac{2}{\e} -\gamma - \frac{{\bar m}^2}{\e} + \left(\frac12 {\bar m}^2 -1\right)\left[\psi\left(\frac12 +\nu\right)+ \psi\left(\frac12-\nu\right) \right] \right] + {\cal O}(\e)\nonumber\\
\label{d4'}
\end{eqnarray}
where $\psi$ stands for the digamma function, and the bar denotes scaling with respect to $H^2$.
Thus in the coincidence limit, we have (under the dimensional regularisation scheme),
\begin{eqnarray}
 &&iG(x,x) = \frac{H^{2-\e}}{2^{3-\e}\pi^{2-\e/2}}\left[\frac{2}{\e} -\gamma - \frac{{\bar m}^2}{\e} + \left(\frac12 {\bar m}^2 -1\right)\left[\psi\left(\nu+\frac12\right)+ \psi\left(\frac12-\nu\right) \right] \right] + {\cal O}(\e)
\label{d4}
\end{eqnarray}
We now define a parameter
\be
s=\frac32 -\nu = \frac32 - \left( \frac94-{\bar m}^2\right)^{1/2}
\label{d5}
\ee
Thus we have 
\begin{eqnarray}
 &&iG(x,x) = \frac{H^{2-\e}}{2^{3-\e}\pi^{2-\e/2}}\left[\frac{2}{\e} -\gamma - \frac{{\bar m}^2}{\e} + \left(\frac12 {\bar m}^2 -1\right)\left[\psi\left(2+s\right)+ \psi\left(-1+s\right) \right] \right] + {\cal O}(\e)\nonumber\\&&
 =\frac{H^{2-\e}}{2^{3-\e}\pi^{2-\e/2}}\left[\frac{2}{\e} -\gamma - \frac{{\bar m}^2}{\e} + \left(\frac12 {\bar m}^2 -1\right)\left(\psi\left(1+s\right)+ \psi\left(1-s\right)+ \frac{2}{1-s}-\frac{1}{s} \right) \right] + {\cal O}(\e),
\label{d60}
\end{eqnarray}
where we have used
$$\psi(1+x)=\psi(x)+\frac{1}{x}$$
\ref{d60} can be renormalised by multiplying it with $m^2/2$ and then using the cosmological constant counterterm, giving  
\begin{eqnarray}
 &&iG(x,x)_{\rm Ren.} = \langle \phi^2 \rangle_{\rm Ren.} 
 =\frac{H^{2}}{2^{3}\pi^{2}}\left(1-\frac12 {\bar m}^2\right)\left[\frac{1}{s}-\frac{2}{1-s}- \left(\psi(1+s)+ \psi(1-s)\right)\right]
\label{d6'}
\end{eqnarray}
In particular, when the field is light i.e. ${\bar m}$ is small, we may expand the digamma function as  
$$\psi(1+s) = -\gamma +\sum_{n=2}^{\infty} (-1)^n \zeta(n)s^{n-1}$$
to have 
\begin{eqnarray}
 &&iG(x,x) = \langle \phi^2\rangle
 =\frac{H^{2-\e}}{2^{3-\e}\pi^{2-\e/2}}\left[\frac{2}{\e} -\gamma - \frac{{\bar m}^2}{\e} + \left(\frac12 {\bar m}^2 -1\right)\left( -2\gamma- 2\sum_{n=1}^{\infty}\zeta(2n+1)s^{2n}+ \frac{2}{1-s}-\frac{1}{s} \right) \right] + {\cal O}(\e)\nonumber\\
\label{d6}
\end{eqnarray}

\bigskip

\noindent
Likewise, the propagator for a fermion field with mass $M$ for small separation  reads, e.g.~\cite{Miao:2012bj}, 
\begin{eqnarray}
 &&iS(x,x') 
 = \frac{\Gamma(1-\e/2)}{2^2\pi^{2-\e/2} (aa')^{3/2-\e/2}}\left[i \slashed{\p}+a M \right]\frac{1}{\Delta x^{2-\e}} + \frac{M H^{2-\e}}{2^{2-\e/2}\pi^{2-\e/2}} \frac{\Gamma \left(\frac{d}{2}+ i{\bar M} \right) \Gamma \left(\frac{d}{2}- i{\bar M} \right)}{\Gamma (1+ i{\bar M})\Gamma (1- i{\bar M})} \left(-\frac{2}{\e}+\gamma-1 \right) \times {\bf I}_{d\times d} +{\cal O}(y) \nonumber\\
\label{d7}
\end{eqnarray}
Hence we have in the dimensional regularisation scheme 
\begin{eqnarray}
 &&  \langle \bar{\psi}\psi\rangle = -{\rm Tr} iS(x,x) 
 = - \frac{M H^{2-\e}}{2^{-\e/2}\pi^{2-\e/2}} \frac{\Gamma \left(\frac{d}{2}+ i{\bar M} \right) \Gamma \left(\frac{d}{2}- i{\bar M} \right)}{\Gamma (1+ i{\bar M})\Gamma (1- i{\bar M})}\left(-\frac{2}{\e}+\gamma -\frac12 \right)\nonumber\\&&
 = - \frac{{\bar M }H^{3-\e}}{2^{-\e/2}\pi^{2-\e/2}}\left(-\frac{2}{\e}+\gamma -\frac12 \right) \left[1-\e +{\bar M}^2  -(1+{\bar M}^2) \left(\psi(1+i{\bar M})+ \psi(1-i{\bar M})\right)\frac{\e}{2} \right] \nonumber\\&&
 = - \frac{{\bar M }H^{3-\e}}{2^{-\e/2}\pi^{2-\e/2}} \left[\left(-\frac{2}{\e}+\frac32+\gamma \right)+ \left(-\frac{2}{\e}-\frac12+\gamma \right){\bar M}^2+ (1+{\bar M}^2) \left(\psi(1+i{\bar M})+ \psi(1-i{\bar M}) \right) \right] 
 \label{d7}
\end{eqnarray}
In particular, when $\bar{M}$ is small, we have 
\begin{eqnarray}
\langle \bar{\psi}\psi\rangle = - \frac{{\bar M }H^{3-\e}}{2^{-\e/2}\pi^{2-\e/2}} \left[\left(-\frac{2}{\e}+\frac32-\gamma \right)+ \left(-\frac{2}{\e}-\frac12-\gamma \right){\bar M}^2- 2(1+{\bar M}^2) \sum_{n=1}^{\infty}\zeta(2n+1)(i\bar{M})^{2n} \right]
\label{d7'}
\end{eqnarray}
If we multiply \ref{d7} by the fermion mass $M$, the corresponding  quantity represents the trace of the free fermion's energy-momentum tensor. Accordingly, it can be renormalised via a cosmological constant counterterm, giving   
\begin{eqnarray}
 &&  \langle \bar{\psi}\psi\rangle_{\rm Ren.}  
 = - \frac{{\bar M }H^{3}}{\pi^{2}} \left[\left(\frac32+\gamma \right)- \left(\frac12-\gamma \right){\bar M}^2+ (1+{\bar M}^2)  \left(\psi(1+i{\bar M})+ \psi(1-i{\bar M}) \right) \right]
\label{d7'}
\end{eqnarray}

\noindent
Note that for non-perturbative computations in the main text, we need to replace the mass terms appearing above by the respective effective dynamical masses.

\section{Local self energy integral with fermion propagators for \ref{y57}}\label{B}
We consider the self energy integral with two fermion propagators appearing in \ref{y57}, and compute {\it only} its local part for our present purpose,                                         
\begin{eqnarray}
&& i \int d^d x'' a''^d {\rm Tr } (iS(x,x'') iS(x'',x)) i  G(x'', x') \nonumber\\&&   = \frac{i}{\pi^{4-\e}}\left\{{\rm Tr} \int d^d x'' \frac{a''^d}{(aa'')^{3-\e}} \left[\frac{\Gamma^2(2-\e/2)}{2^2} \frac{\gamma_{\mu}\gamma_{\nu}\Delta x^{\mu}\Delta x^{\nu}}{\Delta x^{8-2\e}} + \frac{\Gamma^2(1-\e/2) a^2 M^2}{2^4 \Delta x^{4-2\e}}\right] \right\}iG(x'', x') \nonumber\\&&
= \frac{\mu^{-\e} \Gamma(1-\e/2)(1-\e/4)}{2^2 \pi^{2-\e/2}(1-\e)\e} \int d^d x'' \frac{a''^d}{(aa'')^{3-\e}}\left[-\p^2 \delta^d (x-x'') +2 a^2 M^2 \delta^d(x-x'') \right] iG(x'', x')+{\rm non-local~terms} 
\nonumber\\&& =  \frac{\mu^{-\e}  (M^2+H^2) \Gamma(1-\e/2)(1-\e/4)}{2 \pi^{2-\e/2}(1-\e)}\left(\frac{1}{\e}+ \ln a + \frac{\e \ln^2 a}{2}+{\cal O}(\e^2)\right) iG(x,x') +{\rm non-local~terms}
\label{B1}
\end{eqnarray}
where the fermion propagator is given by \ref{d7}.
 The gamma matrices appearing above are flat space, satisfying $[\gamma^{\mu},\gamma^{\nu}]_+= -2\eta^{\mu\nu}\ {\bf I}_{d\times d}$, appropriate for our mostly positive metric signature.  We have also used \ref{A4} in the second equality. \\

 \noindent
For the two loop Yukawa vacuum graph, we need to put $x=x'$ in \ref{B1}, where the coincidence scalar propagator $iG(x,x)$ is given by \ref{d60}. Also, while using in the main text, the mass terms appearing above needs to replaced by the respective effective dynamical masses.

\section{On the difference of results between 2PI renormalisation and the standard 1PI perturbative method}\label{E}

In this appendix 
we wish to sketch very briefly about what happens if we instead attempt to compute the standard perturbative 1PI effective action for the $\phi^4$-Yukawa theory at two loop. First, all the propagators involved here will be tree level, and there is no Schwinger-Dyson  equations containing self energies (\ref{y46}, \ref{y57}). This means, most importantly, there is {\it no} finite loop contribution like $\lambda f_{\rm fin}/2$  in the  mass term of the scalar propagator, and hence we simply take, $m^2_{\rm dyn, eff}=m_0^2+\lambda v^2/2$ in $iG(x,x)$, \ref{y41a1}, \ref{y41a2} and \ref{y57'}.  (The fermion effective mass has already been taken to be, $M_0+gv$, i.e., the usual one for perturbative computations, cf., the discussion below \ref{y52a}). Second, in \ref{y64}, we must have $\delta \lambda_3=0=\delta g_3$.  Accordingly, the constant $C$ is entirely divergent now,  \ref{C-def}. We must also set $k=0$ in \ref{y65} for this perturbative computations, as the scalar mass is now the tree level. Putting everything together, the Yukawa vacuum loop term in \ref{y65} in this case reads
$$ \frac{\mu^{-\e} g^2\Gamma(1-\e/2)(1-\e/4)}{2^2\pi^{2-\e/2}(1-\e)\e}\int a^d d^d x
\left(H^2+(M_0+gv)^2\right)\left( 1+ \e\ln a+ \frac12\e^2\ln^2 a+{\cal O}(\e^3) \right)\left[\left(m_0^2+\frac{\lambda v^2}{2}\right)f_d+ H^2 f'_d + f_{\rm fin}\right]$$
In this perturbative procedure one then adds to the effective action the one loop counterterm contributions, generated by the divergence of the second line of \ref{y64}. After that one fixes the two loop mass and vertex counterterms. However, even after doing that, there remains a non-trivial divergent term in the above expression, explicitly reading,
$$\frac{\mu^{-\e} g^2\Gamma(1-\e/2)(1-\e/4)}{2^2\pi^{2-\e/2}(1-\e)}\int a^d d^d x
\left(H^2+(M_0+gv)^2\right)\ln a\left[\left(m_0^2+\frac{\lambda v^2}{2}\right)f_d+ H^2 f'_d \right]$$
Note that there is no such divergence in flat spacetime ($a=1$).
In order to tackle this, we add with the above the two loop vacuum graph generated by  a {\it finite} quartic vertex counterterm {\it and }  that of the quadratic term containing $\left(H^2+(M_0+gv)^2\right)$, looking explicitly like,
$$\frac{\delta \lambda_{\rm fin}}{2} \int (aa')^d d^x d^d x' \left(H^2+(M_0+gv)^2\right) i \Delta_{++}^2(x,x') i\Delta_{++} (x',x')  $$
 Using now \ref{props2} and \ref{A4}, one can see that the square of the Feynman propagator generates a secular logarithm, and by choosing then $\delta \lambda_{\rm fin}$ appropriately, we can remove the aforementioned divergence. We refer our reader to~\cite{Bhattacharya:2023yhx} and references therein for the standard two loop renormalisation of the $\phi^4$-Yukawa theory in de Sitter. Note in particular that the finite secular logarithm term generated by the Yukawa vacuum loop  is quadratic, opposed to  the non-perturbative result of \ref{y71}.

\bigskip
\bigskip

\end{document}